\documentclass[twocolumn,showpacs]{revtex4}
\usepackage{epsfig}
 \makeindex

 \newcommand\la{\langle}
 \newcommand\ra{\rangle}
 \newcommand\beq{\begin{equation}}
 
 \newcommand\eeq{\end{equation}}
 \newcommand\beqn{\begin{eqnarray}}
 \newcommand\eeqn{\end{eqnarray}}
 \newcommand\GeV{{\rm GeV}}

\def\BA{\begin{eqnarray}}
\def\BE{\begin{equation}}
\def\BF{\begin{figure}[htb]}
\def\BT{\begin{table}[htb]}
\def\EA{\end{eqnarray}}
\def\EE{\end{equation}}
\def\EF{\end{figure}}
\def\ET{\end{table}}

\def\la{\langle}
\def\ra{\rangle}
\def\mb{\,\mbox{mb}}
\def\fm{\,\mbox{fm}}
\def\GeV{\,\mbox{GeV}}

\def\lsim{\mathrel{\rlap{\lower4pt\hbox{\hskip1pt$\sim$}}
    \raise1pt\hbox{$<$}}}         
\def\gsim{\mathrel{\rlap{\lower4pt\hbox{\hskip1pt$\sim$}}
    \raise1pt\hbox{$>$}}}         

\begin{document}


\title
{Production of Polarized Vector Mesons off Nuclei}

\author
{
B.Z. Kopeliovich$^{1,2}$, J. Nemchik$^{3}$ and Ivan~Schmidt$^1$
}
\affiliation
{
$^{1}${ Departamento de F\'isica y Centro de Estudios Subat\'omicos, 
Universidad T\'ecnica Federico Santa Mar\'ia, Valpara\'iso, Chile}\\
$^{2}${Joint Institute for Nuclear Research, Dubna, 141980 Moscow
Region, Russia}  \\
$^{3}${Institute of Experimental Physics SAS, Watsonova 47,
04001 Kosice, Slovakia}
}


\begin{abstract}

Using the light-cone QCD dipole formalism we investigate
manifestations of color transparency (CT) and coherence length (CL)
effects in electroproduction of longitudinally (L) and transversally
(T) polarized vector mesons. Motivated by forthcoming data from the
HERMES experiment we predict both the $A$ and $Q^2$ dependence of
the $L/T$- ratios, for $\rho^0$ mesons produced coherently and
incoherently off nuclei. For an incoherent reaction the CT and CL
effects add up and result in a monotonic $A$ dependence of the
$L/T$-ratio at different values of $Q^2$. On the contrary, for a
coherent process the contraction of the CL with $Q^2$ causes an
effect opposite to that of CT and we expect quite a nontrivial $A$
dependence, especially at $Q^2\gg m_V^2$.

\end{abstract}


\maketitle

%
\section{Introduction}
\label{intro}
%

Electroproduction of vector mesons has been intensively studied
during the last three decades.  Numerous fixed target experiments
have provided high quality data: the OMEGA \cite{omega} and NMC 
\cite{nmc} experiments at CERN, CHIO  
\cite{chio} and E665 experiments at Fermilab \cite{e665}, 
etc. \cite{fix-target-all}. In particular, the E665
collaboration \cite{e665-rho} observed for the first time a
manifestation of color transparency (CT) 
\cite{zkl,bbgg}
in vector meson production
on nuclear targets. The data confirmed the
predictions for CT presented in Ref.~\cite{knnz94}.

Moreover, important dynamical information on vector meson
electroproduction, in a wide range of  $Q^2$ and energies and for
different photon polarizations, transversal (T) and longitudinal
(L), was provided by experiments performed by both the ZEUS
\cite{zeus-all} and H1 \cite{h1-all} collaborations at HERA. In
particular it was found that the longitudinal-to-transverse  ratio,
$R_{LT}$, for exclusive electroproduction of $\rho^0$ mesons, rises
with $Q^2$, but has a weak energy dependence at fixed $Q^2$
\cite{zeus-lt1,zeus-lt2,zeus-ro-prel,zeus-fi-prel,h1-lt1,h1-lt2,h1-ro-prel,
h1-lt3,hermes-lt,hermes-lts,nmc,e665}. Some correlation between the
energy and $Q^2$ dependence was detected in the ZEUS experiment
\cite{zeus-lt1}:
the energy dependence
of the ratio $R_{LT}$ is stronger at larger $Q^2$. 

These observations can be understood within the dipole approach \cite{bfkl94}.
The shrinkage of the $\bar qq$ component of the photon with $Q^2$ and the
small-size behavior of the dipole cross section \cite{zkl}, $\sigma_{\bar qq}(r)
\sim r^2$
\footnote{$\sigma_{\bar qq}(r)$ is the cross section for the interaction with a
nucleon of the $\bar qq$ fluctuations of the photon having transverse separations
$\vec{r}$},
 lead to so called {\sl scanning effect} \cite{knnz93,knnz94,bfkl94}. Namely, the
vector meson production amplitude is dominated by the contribution dipole sizes
of the order of $r \sim r_S$, 
 %
 \beq
r_S \approx \frac{Y}{\sqrt{Q^2 + m_V^2}}\ ,
 \label{10}
 \eeq
 %
 where the product of the photon wave function and the dipole cross section forms
a sharp peak. Varying $Q^2$ and the mass of vector meson $m_V$ one can study the
transition from the nonperturbative region of large $r_S$ to the perturbative
region of very small $r_S\ll R_V$, where $R_V$ is the radius of vector meson.

Factor $Y$ in Eq.(\ref{10}) was evaluated in \cite{bfkl94} at $Y\approx 6$.
However, this estimate made use of a nonrelativistic approximation which is
reasonable for charmonium and is rather accurate for bottonium production.
Moreover, in general case the parameter $Y$ depends on polarization and increases
slowly with $Q^2$. The $Q^2$ dependence of $Y_{T,L}$ is related to so called
aligned-jet configurations of the $\bar qq$ configurations when $q$ or $\bar q$
carry almost the whole momentum of the photon. Since these end-point
configurations in a longitudinally polarized photons are suppressed, one should
expect $Y_{L} < Y_{T}$. In another words, the production amplitude of
longitudinal vector mesons scans their wave function at smaller transverse sizes.
According to Eq.~(\ref{10}) higher $Q^2$ results in a smaller transverse size of
the color $\bar qq$ dipole, i.e. in a smaller $r_S$. Stronger energy dependence
of the dipole cross section $\sigma_{\bar qq}(r,s)$ at smaller dipole size causes
a weak energy dependence of the ratio $R_{LT}$ (see also \cite{jan97}). However
large errors of available data do not allow to see clearly this energy
dependence.

Eq.~(\ref{10}) shows that one can reach small perturvative scanning radius only
at very large scale $Q^2\gg m_V^2$.  
%
%
%
%

There are many experimental and theoretical evidences \cite{spots,drops} that a
semihard scale of a nonperturbative origin exists in QCD \cite{kst2}. Namely, the
mean transverse distance of gluon propagation is small, of the order of
$r_0\sim0.3\fm$. In order to rely on pQCD calculations one should make the
scanning radius smaller than $r_0$, i.e.
 %
 \BE
 (Q^2 + m_V^2)\gsim Q_{pQCD}^2 = \frac{Y^2}{r_0^2}\, .
 \label{16}
 \EE
 %

As was discussed in Refs.~\cite{knst-01,n-02,n-03}, nuclear targets represent
unique analyzers of the dynamics of vector meson production. They allow to study
other phenomena such as color transparency (CT), coherence length (CL) effects,
gluon shadowing (GS), etc. These effects were studied in
\cite{knst-01} for coherent and incoherent electroproduction of vector mesons,
and within a quantum-mechanical description of the $\bar qq$ evolution, based on
the light-cone (LC) Green function technique \cite{kz-91}. The same LC Green
function formalism has been applied also for Drell-Yan production in
proton-nucleus and nucleus-nucleus interactions \cite{krtj-03}, and for nuclear
shadowing in DIS \cite{krt1,krt2}.

Data for vector meson production off nuclei are usually presented
in the form of the so called nuclear transparency, defined as a ratio:
%
\beq
Tr_A = \frac{\sigma_{\gamma^*A\rightarrow VX}}
                  {A~\sigma_{\gamma^*N\rightarrow VX}}
\label{15}
\eeq
%
for the diffractive incoherent (quasielasic) production of vector mesons,
$\gamma^*A\rightarrow VX$, where one sums over all final states of the
target nucleus except those which contain particle (pion) creation.

If electroproduction of a vector meson leaves the target intact, the process
$\gamma^*A\rightarrow VA$ is usually called coherent or elastic. For this process
one can formally define the nuclear transparency in the same way, Eq.~(\ref{15}),
however, the coherent production cross section $\sigma_{\gamma^*A\rightarrow
VA}^{coh}$ has a form different from the incoherent cross section
$\sigma_{\gamma^*A\rightarrow VX}^{inc}$, as will be seen below in
Sections~\ref{vm-incoh} and \ref{vm-coh} (see Eqs.~(\ref{520}) and (\ref{550})).

There are two time scales which control the dynamics of vector meson
production \cite{knst-01}. The first time scale is called formation
time, and is connected with the phenomenon called color
transparency. This effect comes from QCD and has been studied
intensively for almost two decades. The second time scale is known
as the coherence time and is connected with quantum coherence
effects. Both phenomena cause nuclear suppression.

The phenomenon of CT can be treated either in the hadronic or in the
quark basis. The former approach leads to Gribov's inelastic
corrections \cite{gribov}, while the latter manifests itself as a
result of color screening \cite{zkl,bbgg}. Although these two
approaches are complementary, the quark-gluon interpretation is more
intuitive and straightforward, coming from the fact that colorless
hadrons can interact only because color is distributed inside them.
If the hadron transverse size $r$ tends to zero then the interaction
cross section $\sigma_{\bar qq}(r)$ vanishes as $r^2$ \cite{zkl}. As
a result the nuclear medium is more transparent for smaller
transverse size hadrons. Besides, this fact naturally explains the
correlation between the cross sections of hadrons and their sizes
\cite{gs,hp,p}.

Diffractive electroproduction of vector mesons off nuclei is one of
the most effective processes for studying CT. According to
Eq.~(\ref{10}), in this case a photon of high virtuality $Q^2 \gg
m_V^2$ is expected to produce a pair with a small transverse
separation $\sim 1/Q^2$ ~\footnote{In fact, the situation is
somewhat more complicated. For very asymmetric pairs, when the $q$
or $\bar q$ carry almost the whole photon momentum, the pair can
have a large separation, see Sect.~\ref{lcc}}. Then CT manifests
itself as a vanishing absorption of the small sized colorless $\bar
qq$ wave packet during propagation through the nucleus. The
dynamical evolution of this small size $\bar qq$ pair into a normal
size vector meson is controlled by the time scale called formation
time. Due to uncertainty principle, one needs a time interval to
resolve different levels $V$ (the ground state) or $V'$ (the next
excited state) in the final state. In the rest frame of the nucleus
this formation time is Lorentz dilated,
%
 \beq
t_f = \frac{2\,\nu}
{\left.m_V^\prime\right.^2 - m_V^2}\ ,
\label{20}
 \eeq
%
where $\nu$ is the photon energy. A rigorous quantum-mechanical
description of the pair evolution was suggested in \cite{kz-91} and
is based on the nonrelativistic light-cone Green function technique.
A complementary description of the same process in the hadronic
basis is presented in \cite{hk-97}.

Another phenomenon known to cause nuclear suppression is quantum
coherence, which results from the destructive interference of
amplitudes for which the interaction takes place on different bound
nucleons. It is controlled by the distance from the production to
the absorption point when the pointlike photon becomes the
hadronlike $\bar qq$ pair, and which may be also interpreted as the
lifetime of $\bar qq$ fluctuation, thus providing the time scale
which controls shadowing. Again, it can be estimated relying on the
uncertainty principle and Lorentz time dilation as,
%
 \beq
t_c = \frac{2\,\nu}{Q^2 + m_V^2}\ .
\label{30}
 \eeq
%
This is usually called coherence time, but we also will use the term
coherence length, since light-cone kinematics is assumed, $l_c=t_c$
(similarly, for formation length $l_f=t_f$). The CL is related to
the longitudinal momentum transfer  in $\gamma^*\,N \to V\,N$ as
$q_c=1/l_c$, which controls the interference of the production
amplitudes from different nucleons.

Since the exclusive production of vector mesons at high energies is
controlled by small-$x_{Bj}$ physics, gluon shadowing becomes an
important issue \cite{knst-01}. In fact, GS suppresses
electroproduction of vector mesons. Although it has been shown
\cite{ikth-02} that for electroproduction of charmonia off nuclei GS
starts to be important at center-of-mass energies $\sqrt(s) \geq
30-60\GeV$, the same does not happen for electroproduction of light
vector mesons \cite{knst-01}, where GS starts to be effective at
smaller energy values $\sqrt(s) \geq 7-30\GeV$. Nevertheless, GS in
the HERMES kinematical range discussed in the present paper is
negligible and does not need to be included in calculations.

In electroproduction of vector mesons off nuclei one needs to
disentangle CT (absorption) and CL (shadowing) as the two sources of
nuclear suppression. These effects can be associated with final and
initial state interactions, respectively. A detailed analysis of the
CT and CL effects in electroproduction of vector mesons off nuclei
showed \cite{knst-01}, for the example of vector dominance model
(VDM), that one can easily identify the difference of the nuclear
suppression corresponding to absorption and shadowing,
in the two limiting cases:\\
i.) The limit of $l_c$ shorter than the mean internucleon spacing
$\sim 2\,fm$. In this case only final state absorption matters. The
ratio of the quasielastic (or incoherent) $\gamma^*\, A \to V\,X$
and $\gamma^*\, N \to V\,X$ cross sections, usually called nuclear
transparency, reads \cite{kz-91},
%
\begin{widetext}
 \BA
Tr_A^{inc}\Bigr|_{l_c\ll R_A} &\equiv&
\frac{\sigma_V^{\gamma^*A}}
{A\,\sigma_V^{\gamma^*N}} =
\frac{1}{A}
\,\int d^2b\,
\int\limits_{-\infty}^{\infty}
dz\,\rho_A(b,z)\,
\exp\left[-\sigma^{VN}_{in}
\int\limits_z^{\infty} dz'\,
\rho_A(b,z')\right]\nonumber\\
&=&
\frac{1}{A\,\sigma^{VN}_{in}}\,
\int d^2b\,\left\{1 -
\exp\Bigl[-\sigma^{VN}_{in}\,T_A(b)\Bigr]\right\}=
\frac{\sigma^{VA}_{in}}{A\,\sigma^{VN}_{in}}\ .
\label{40}
 \EA
\end{widetext}
%
Here $z$ is the longitudinal coordinate and $\vec{b}$ the impact
parameter of the production point of vector meson. In (\ref{40})
$\rho_A(b,z)$ is the nuclear density and $\sigma_{in}^{VN}$ is the
inelastic $VN$ cross section.

ii.) The limit of long $l_c$, where the expression for the nuclear
transparency takes a different form,
%
 \beq
Tr_A^{inc}\Bigr|_{l_c\gg R_A} =
\int d^2b\,T_A(b)\,
\exp\left[-\sigma^{VN}_{in}\,T_A(b)\right]\ ,
\label{50}
 \eeq
%
and where we assume $\sigma^{VN}_{el} \ll \sigma^{VN}_{in}$ for the
sake of simplicity. Here $T_A(b)$ is the nuclear thickness function
%
 \beq
T_A(b) = \int\limits_{-\infty}^{\infty} dz\,\rho_A(b,z)\ .
\label{60}
 \eeq
%

The exact expression which interpolates between the two regimes
(\ref{40}) and (\ref{50}) can be found in Ref.~\cite{hkn}.

The problem of CT-CL separation is different depending on the mass
of the produced vector meson. In the production of light vector
mesons ($\rho^0$, $\Phi^0$) \cite{knst-01} the coherence length is
larger or comparable with the formation length, $l_c\gsim l_f$,
starting from the photoproduction limit up to $Q^2 \sim (1\div
2)\GeV^2$. For charmonium production, however, there is a strong
inequality $l_c < l_f$ independent of $Q^2$ and $\nu$
\cite{n-02,n-03}, which therefore leads to a different scenario of
CT-CL mixing.

Recently new HERMES data \cite{hermes-a1,hermes-a2} for diffractive exclusive
electroproduction of $\rho^0$ mesons on nitrogen target have been gradually
become available. At the beginning the data were presented as a dependence of
nuclear transparencies (\ref{15}) on coherence length (\ref{30}). The data for
incoherent $\rho^0$ production decrease with $l_c$, as expected from the effects
of initial state interactions. On the other hand, the nuclear transparency for
coherent $\rho^0$ production increases with coherence length, as expected from
the effects of the nuclear form factor \cite{knst-01}. However, each $l_c$-bin of
the data contains different values of $\nu$ and $Q^2$, i.e. there are different
contributions from both effects, CT and CL. For this reason the $l_c$ behavior of
nuclear transparency does not allow to study separately the CT and CL effects.
Therefore it was proposed in \cite{hk-97,knst-01} that CT can be separately
studied, eliminating the effect of CL from the data on the $Q^2$ dependence of
nuclear transparency, in a way which keeps $l_c = const$. According to this
prescription, the HERMES data \cite{hermes-a2} were later presented as the $Q^2$
dependence of nuclear transparency, albeit at different fixed values of $l_c$.
Then the rise of $Tr_A^{inc}$ and $Tr_A^{coh}$ with $Q^2$ represents a signature
of CT. The HERMES data \cite{hermes-a2} are in a good agreement with the
predictions from Ref.~\cite{knst-01}.

New HERMES data on neon and krypton targets should be presented
soon, which will allow to verify further the predictions for CT from
\cite{knst-01}. Besides, gradually increasing the statistics of the
HERMES data should allow to obtain also the results at different
polarizations L and T, and this gives the interesting possibility of
studying the polarization dependence of the CT and CL effects, in
both coherent and incoherent production of vector mesons. The data
are usually presented as the $L/T$-ratio $R_{LT}^A$ of the
corresponding nuclear cross sections. Knowing the nucleon
$L/T$-ratio $R_{LT}$ one can define the nuclear modification factor
as
%
\BE
f (s,Q^2,A) = \frac{R_{LT}^A}{R_{LT}}\, .
\label{65}
\EE
%
for both coherent and incoherent processes. The nuclear modification
factor Eq.~(\ref{65}) represents a modification of the nucleon
$L/T$- ratio, given by a nuclear environment. Its deviation from
unity allows to obtain information about a possible different onset
of CT and CL effects in the production of L and T vector mesons.
Therefore an exploratory study of the $Q^2$ and $A$ dependence of
the factors $f_{inc}$ and $f_{coh}$ gives an alternative way for the
investigation of CT and CL effects in coherent and incoherent
production of vector mesons, at different polarizations L and T.
This is the main goal of the present paper.

The paper is organized as follows. In Sect.~\ref{lcc} we present the
light-cone approach to diffractive electroproduction of vector
mesons in the rest frame of the nucleon target. Here we also
describe the individual ingredients contained in the production
amplitude: (i) the dipole cross section, (ii) the LC wave function
for a quark-antiquark fluctuation of the virtual photon, and (iii)
the LC wave function of the vector meson.

In the next Sect.~\ref{data-N} we calculate the nucleon $L/T$-ratio
$R_{LT}$ of the cross sections for exclusive electroproduction of L
and T polarized $\rho^0$, $\Phi^0$ and charmonia. The model
calculations reproduce quite well the available data for the $Q^2$
dependence of $R_{LT}$. This is an important test of the model
because $R_{LT}$ is included in the calculations of the nuclear
$L/T$-ratio.

Sect.~\ref{vm-incoh} is devoted to the incoherent production of
vector mesons off nuclei. First, in Section~\ref{intro-inc} we
define the different transparency ratios, and in \ref{lc-inc}, we
briefly describe the formalism based on the LC Green function
technique. In Sect.~\ref{regimes} we analyze different regimes of
incoherent production of vector mesons, depending on the magnitude
of the coherence length. Then in Sect.~\ref{rlt-inc} we present a
discussion about the $A$ and $Q^2$ behavior of the nuclear
$L/T$-ratio in the limit of long coherence length $l_c\gg R_A$,
because then the corresponding formulae and theoretical treatment
get simplified with respect to the general case $l_c\sim R_A$, where
there is a strong CT-CL mixing. Here we present also the model
predictions for the nuclear modification factor $f_{inc}$ and
nuclear $L/T$- ratio. Finally we study the general case when there
is no restriction on the coherence length. The numerical
calculations of Sect.~\ref{incoh-data} produce the prediction for
the $L/T$- ratio of nuclear cross sections, for production of L and
T polarized vector mesons, as a function of the mass number $A$ at
different fixed values of $\la Q^2\ra$ corresponding to the HERMES
kinematics. We find a monotonic $A$ dependence of this ratio. We
discuss why this $A$ behavior of $R_{LT}^A(inc)$ only weakly changes
with $Q^2$. Following the prescription of Refs.~\cite{hk-97,knst-01}
we investigate also how a clear signal of CT effects manifests
itself separately in the production of L and T polarized vector
mesons. We present the model predictions for the $Q^2$ dependence of
$f_{inc}$, at different fixed values of the coherence length. Such a
polarization dependence of CT effects can be analyzed by the HERMES
collaboration and in the experiments at JLab.

Coherent production of vector mesons off nuclei leaving the nucleus
intact is studied in Sect.~\ref{vm-coh}. The formalism, with an
emphasis on the nuclear $L/T$- ratio, is described in
Sect.~\ref{lc-coh}. Then, just as for incoherent production, we
analyze in Sect.~\ref{rlt-coh} the $A$ and $Q^2$ behavior of the
nuclear $L/T$-ratio, in the limit of long coherence length. Here we
present also the corresponding model predictions for the nuclear
modification factor and the $L/T$-ratio. The general case with no
restriction on the coherence length is analyzed in
Sect.~\ref{coh-data}. Contrary to incoherent vector meson
production, here we find, at medium and large values of $Q^2$ (when
$l_c\lsim R_A$, where $R_A$ is the nuclear radius), a nonmonotonic
$A$ dependence of the nuclear $L/T$- ratio. This nontrivial and
anomalous $A$ dependence of $R_{LT}^A(coh)$ is even more complicated
at larger values of $Q^2$, as a result of a stronger interplay of CT
and CL effects. We find also a different manifestation of the net CT
effects in the production of L and T polarized vector mesons, by
performing the predictions at fixed values of the coherence length.

The gluon shadowing starts to manifest itself at $\sqrt{s}\geq
7-30\GeV$ and is not significant in the HERMES energy range studied
in the present paper. Therefore it is not included in the
calculations.

The results of the paper are summarized and discussed in
Sect.~\ref{conclusions}. The important conclusion of a nontrivial
$A$ dependence of the coherent nuclear $L/T$-ratio for the expected
new HERMES data and for the future planned experiments is stressed.

%
%
\section{Color dipole phenomenology for elastic
electroproduction
of vector mesons \boldmath$\gamma^{*}N\to V~N$}\label{lcc}
%
%

The LC dipole approach for elastic electroproduction $\gamma^{*}N\to
V~N$ was already used in Ref.~\cite{hikt-00} to study the exclusive
photo and electroproduction of charmonia, and in Ref.~\cite{knst-01}
for elastic virtual photoproduction of light vector mesons $\rho^0$
and $\Phi^0$ (for a review see also \cite{ins-05}). Therefore, we
present only a short review of this LC phenomenology, with the main
emphasis of looking at the effects of the different polarizations L
and T. In this approach a diffractive process is treated as elastic
scattering of a $\bar qq$ fluctuation of the incident particle.  The
elastic amplitude is given by convolution of the universal flavor
independent dipole cross section for the $\bar qq$ interaction with
a nucleon, $\sigma_{\bar qq}$, \cite{zkl} and the initial and final
wave functions. For the exclusive photo- or electroproduction of
vector mesons $\gamma^{*}N\to V~N$ the forward production amplitude
is represented in the following form
%
\begin{widetext}
 \BE
{\cal M}_{\gamma^*N\rightarrow VN}(s,Q^{2})\, =\,
\langle\, V |\,\sigma_{\bar qq}({\vec{r}},s)\,|\gamma^{*}\,\rangle\, =\,
\int\limits_{0}^{1} d\alpha \int d^{2}{{r}}\,\,
\Psi_{V}^{*}({\vec{r}},\alpha)\,
\sigma_{\bar qq}({\vec{r}},s)\,
\Psi_{\gamma^{*}}({\vec{r}},\alpha,Q^2)\,
\label{120}
 \EE
\end{widetext}
%
 with the normalization
%
 \beq
\left.\frac{d\sigma(\gamma^*N\to VN)}{dt}\right|_{t=0} =
\frac{|{\cal M}_{\gamma^*N\to VN}(s,Q^2)|^{2}}{16\,\pi}.
\label{125}
 \eeq
%

There are three ingredients contributing to the amplitude
(\ref{120}):

(i) The dipole cross section $\sigma_{\bar qq}({\vec{r}},s)$, which
depends on the $\bar qq$ transverse separation $\vec{r}$ and the
c.m. energy squared $s$.

(ii) The LC  wave function of the photon
$\Psi_{\gamma^{*}}({\vec{r}},\alpha,Q^2)$, which depends also on the
photon virtuality $Q^2$ and the relative share $\alpha$ of the
photon momentum carried by the quark.

(iii) The LC wave function $\Psi_{V}({\vec{r}},\alpha)$ of the
vector meson.

Notice that in the LC formalism the photon and meson wave functions
contain also higher Fock states $|\bar qq\ra$, $|\bar qqG\ra$,
$|\bar qq2G\ra$, etc. The effect of higher Fock states are
implicitly incorporated into the energy dependence of the dipole
cross section $\sigma_{\bar qq}(\vec{r},s)$, as is given in
Eq.~(\ref{120}).

\subsection{Dipole Cross Section}

The dipole cross section $\sigma_{\bar qq}(\vec r,s)$ represents the
interaction of a $\bar qq$ dipole of transverse separation $\vec r$
with a nucleon \cite{zkl}. It is a flavor independent universal
function of $\vec{r}$ and energy, and allows to describe in a
uniform way various high energy processes. It is known to vanish
quadratically $\sigma_{\bar qq}(r,s)\propto r^2$ as $r\rightarrow
0$, due to color screening (CT property). The dipole cross section
cannot be predicted reliably because of poorly known higher order
pQCD corrections and nonperturbative effects. A detailed discussion
about the dipole cross section in connection with production of
vector mesons is presented in \cite{knst-01}.

There are two popular parametrizations of $\sigma_{\bar qq}(\vec
r,s)$. The first one, suggested in \cite{gbw}, reflects the fact
that at small separations the dipole cross section should be a
function of $r$ and $x_{Bj}\sim 1/(r^2s)$, in order to reproduce
Bjorken scaling. It describes well data for deep-inelastic
scattering (DIS) at small $x_{Bj}$ and medium and high $Q^2$.
However, at small $Q^2$ it cannot be correct since it predicts
energy independent hadronic cross sections. Besides, $x_{Bj}$ is not
any more the proper variable at small $Q^2$ and should be replaced
by energy. This defect is removed by the second parametrization
proposed in \cite{kst2}, which is similar to the first one
\cite{gbw}, but contains an explicit dependence on energy, and it is
valid down to the limit of real photoproduction. Since we will
consider HERMES data with a typical kinematical region of the photon
energy, $5 <\nu < 24\GeV$, and virtuality, $0.8 < Q^2 < 5\GeV^2$, we
choose the second parametrization, which has the following form:
%
 \BA
  \sigma_{\bar qq}(r,s) &=& \sigma_0(s)
\left[1 - e^{- r^2/r_{0}^2(s)}
  \right]\ ,
  \label{130}
 \EA
%
where
%
 \BE
  \sigma_0(s) = \sigma^{\pi p}_{tot}(s)
  \left[1+\frac38 \frac{r_{0}^2(s)}
{\left<r^2_{ch}\right>}\right]\mb\ \\
  \label{140}
 \EE
%
and
%
 \BE
  r_0(s)   = 0.88 \left(\frac{s}{s_0}\right)^{\!\!-0.14}  \fm\ .
  \label{150}
 \EE
%
 Here $\left<r^2_{ch}\right>=0.44\fm^2$ is the mean pion charge radius
squared, and $s_0 = 1000\GeV^2$.  The cross section $\sigma^{\pi p}_{tot}(s)$
was fitted to data in \cite{dl,pdt} and reads
%
 \BE
\sigma^{\pi p}_{tot}(s) =
23.6\,\left(\frac{s}{s_0}\right)^{\!\!0.079}
+ 1.425\,\left(\frac{s}{s_0}\right)^{\!\!-0.45}\,\mb\ .
\label{145}
 \EE
%
This represents the Pomeron and Reggeon parts, corresponding to
exchange of gluons and $\bar qq$, respectively. A detailed
description of the incorporation of Reggeons into the LC dipole
formalism can be found in Ref.~\cite{knst-01}.

The dipole cross section presented in Eqs.(\ref{130}) -- (\ref{145}) provides the
imaginary part of the elastic amplitude.  It is known, however, that the
energy dependence of the total cross section generates also a real part
\cite{bronzan},
%
 \beq
\sigma_{\bar qq}(r,s) \Rightarrow
\left(1-i\,\frac{\pi}{2}\,\frac{\partial}
{\partial\ln(s)}\right)\,
\sigma_{\bar qq}(r,s)\, .
\label{real-part}
 \eeq
%
 Therefore the energy dependence of the dipole cross section given by Eq.~(\ref{130}), which
is rather steep at small $r$, leads to a large real part which
should not be neglected.

\subsection{The $\bar qq$ wave function of the photon}

The perturbative distribution amplitude (``wave function'') of the
$\bar qq$ Fock component of the photon has the following form, for T
and L polarized photons \cite{lc,bks-71,nz-91},
%
 \BE
\Psi_{\bar qq}^{T,L}({\vec{r}},\alpha) =
\frac{\sqrt{N_{C}\,\alpha_{em}}}{2\,\pi}\,\,
Z_{q}\,\bar{\chi}\,\hat{O}^{T,L}\,\chi\,
K_{0}(\epsilon\,r)
\label{70}
 \EE
%
 where $\chi$ and $\bar{\chi}$ are the spinors of the quark and
antiquark, respectively, $Z_{q}$ is the quark charge, with $Z_{q} =
1/\sqrt{2}, 1/3\sqrt{2}, 1/3, 2/3$ and $1/3$ for $\rho^0$,
$\omega^0$, $\Phi^0$, $J/\Psi$ and $\Upsilon$ production
respectively, $N_{C} = 3$ is the number of colors. and
$K_{0}(\epsilon r)$ is a modified Bessel function with
%
 \BE
\epsilon^{2} =
\alpha\,(1-\alpha)\,Q^{2} + m_{q}^{2}\ .
\label{80}
 \EE
%
Here $m_{q}$ is the quark mass, and $\alpha$ is the fraction of the
LC momentum of the photon carried by the quark. The operators
$\widehat{O}^{T,L}$ are given by
%
 \BE
\widehat{O}^{T} = m_{q}\,\,\vec{\sigma}\cdot\vec{e} +
i\,(1-2\alpha)\,(\vec{\sigma}\cdot\vec{n})\,
(\vec{e}\cdot\vec{\nabla}_r) + (\vec{\sigma}\times
\vec{e})\cdot\vec{\nabla}_r\ ,
 \label{90}
 \EE
%
 \BE
\widehat{O}^{L} =
2\,Q\,\alpha (1 - \alpha)\,(\vec{\sigma}\cdot\vec{n})\ .
\label{100}
 \EE
%
 Here $\vec\nabla_r$ acts on the transverse coordinate $\vec r$,
$\vec{e}$ is the polarization vector of the photon, $\vec{n}$ is a
unit vector parallel to the photon momentum and $\vec{\sigma}$ is
the three-vector of the Pauli spin matrices.

The transverse separation of the $\bar qq$ pair contains an explicit
$\alpha$ dependence and can be written using the expression for the
scanning radius, Eq.~(\ref{10}), as
%
 \BA
 r_{\bar qq} &\sim& \frac{1}{\epsilon} =
\frac{1}{\sqrt{Q^{2}\,\alpha\,(1-\alpha) + m_{q}^{2}}}
\nonumber \\
&\sim&
\frac{r_S}{3} =
\frac{\tilde Y}{\sqrt{Q^2 + m_V^2}}\, ,
\label{212}
 \EA
%
where $\tilde Y = Y/3$. For very asymmetric $\bar qq$ pairs the LC
variable $\alpha$ or $(1-\alpha) \lsim m_q^2/Q^2$. Consequently, the
transverse separation $r_{\bar qq} \sim \tilde 1/m_q$ and the
scanning radius $r_S$ become large.
However, this is not the case of charmonium and bottonium production
because of the large quark masses $m_c = 1.5\GeV$ and $m_b =
5.0\GeV$, respectively. Therefore in this later case it is
straightforward to include nonperturbative interaction effects
between $q$ and $\bar q$. In the production of light vector mesons
there are two ways how to fix the problem of a huge $\bar qq$
transverse separation. One can introduce an effective quark mass
$m_q\approx \Lambda_{QCD}$, which should represent the
nonperturbative interaction effects between the $q$ and $\bar q$,
while the other way is to introduce this interaction explicitly. We
use the second possibility, with the corresponding phenomenology
based on the LC Green function approach developed in
Ref.~\cite{kst2}.

The Green function $G_{\bar qq}(z_1,\vec r_1;z_2,\vec r_2)$
describes the propagation of an interacting $\bar qq$ pair between
points with longitudinal coordinates $z_{1}$ and $z_{2}$, and with
initial and final separations $\vec r_1$ and $\vec r_2$. This Green
function satisfies the two-dimensional Schr\"odinger equation,
%
\BA
i\frac{d}{dz_2}\,G_{\bar qq}(z_1,\vec r_1;z_2,\vec r_2) =\,\,\,\,\,\,\,\,\,\,
~~~~~~~~~~~~
\nonumber \\
\left\{\frac{\epsilon^{2} - \Delta_{r_{2}}}{2\,\nu\,\alpha\,(1-\alpha)}
+ V_{\bar qq}(z_2,\vec r_2,\alpha)\,\right\}
G_{\bar qq}(z_1,\vec r_1;z_2,\vec r_2)\ .
\label{250}
\EA
%
 Here $\nu$ is the photon energy, and the Laplacian $\Delta_{r}$ acts on
the coordinate $r$.

The imaginary part of the LC potential $V_{\bar qq}(z_2,\vec
r_2,\alpha)$ in (\ref{250}) is responsible for the attenuation of
the $\bar qq$ in the medium, while the real part represents the
interaction between the $q$ and $\bar{q}$. This potential is
supposed to provide the correct LC wave functions of the vector
mesons. For the sake of simplicity we use  the oscillator form of
the potential,
%
 \BE
{\rm Re}\,V_{\bar qq}(z_2,\vec r_{2},\alpha) =
\frac{a^4(\alpha)\,\vec r_{2}\,^2}
{2\,\nu\,\alpha(1-\alpha)}\ ,
\label{260}
 \EE
%
 which leads to a Gaussian $r$-dependence of the LC wave function of the
meson ground state.  The shape of the function $a(\alpha)$ will be
discussed below.

 In this case equation (\ref{250}) has an analytical solution, leading
to an explicit form of harmonic oscillator Green function \cite{fg},
%
\begin{widetext}
 \BA
G_{\bar qq}(z_1,\vec r_1;z_2,\vec r_2) =
\frac{a^2(\alpha)}{2\;\pi\;i\;
{\rm sin}(\omega\,\Delta z)}\, {\rm exp}
\left\{\frac{i\,a^2(\alpha)}{{\rm sin}(\omega\,\Delta z)}\,
\Bigl[(r_1^2+r_2^2)\,{\rm cos}(\omega \;\Delta z) -
2\;\vec r_1\cdot\vec r_2\Bigr]\right\}
{\rm exp}\left[-
\frac{i\,\epsilon^{2}\,\Delta z}
{2\,\nu\,\alpha\,(1-\alpha)}\right] \ ,
\label{270}
 \EA
\end{widetext}
%
where $\Delta z=z_2-z_1$ and
%
 \BE \omega = \frac{a^2(\alpha)}{\nu\;\alpha(1-\alpha)}\ .
\label{280}
 \EE
%
 The boundary condition is $G_{\bar
qq}(z_1,\vec r_1;z_2,\vec r_2)|_{z_2=z_1}= \delta^2(\vec r_1-\vec
r_2)$.

The probability amplitude to find the $\bar qq$ fluctuation of a
photon at the point $z_2$ with separation $\vec r$ is given by an
integral over the point $z_1$ where the $\bar qq$ is created by the
photon with initial separation zero,
%
 \BA
\Psi^{T,L}_{\bar qq}(\vec r,\alpha)=
\frac{i\,Z_q\sqrt{\alpha_{em}}}
{4\pi\,\nu\,\alpha(1-\alpha)}\times
\nonumber \\
\int\limits_{-\infty}^{z_2}dz_1\,
\Bigl(\bar\chi\;\widehat O^{T,L}\chi\Bigr)\,
G_{\bar qq}(z_1,\vec r_1;z_2,\vec r)
\Bigr|_{r_1=0}\ .
\label{290}
 \EA
%
 The operators $\widehat O^{T,L}$ are defined by Eqs.~(\ref{90}) and
(\ref{100}). Here they act on the coordinate $\vec r_1$.

If we write the transverse part  as
%
 \BA
\bar\chi\;\widehat O^{T}\chi\,&=&\,
\bar\chi\,m_{q}\,\vec{\sigma}\cdot\vec{e}\,\chi\, +\,
\bar\chi\,\bigl [i\,(1-2\alpha)\,(\vec{\sigma}\cdot\vec{n})\,
\vec{e}
\nonumber\\
&&+ (\vec{\sigma}\times
\vec{e}\,)\bigr ]\chi\,\cdot\vec{\nabla}_r\ ,
\nonumber\\
&=&
E + \vec F\cdot\vec\nabla_{r}\ ,
\label{300}
 \EA
%
then the distribution functions read,
%
 \BA
\!\!\!\Psi^{T}_{\bar qq}(\vec r,\alpha) =
Z_q\sqrt{\alpha_{em}}\,
\left[E\,\Phi_0(\epsilon,r,\lambda)
+ \vec F\cdot\vec\Phi_1(\epsilon,r,\lambda)\right],
\nonumber\\
\label{310}
 \EA
%
 \BA
\!\!\!\Psi^{L}_{\bar qq}(\vec r,\alpha) =
2\,Z_q\sqrt{\alpha_{em}}\,
Q\,\alpha(1-\alpha)\,
\bar\chi\,\vec\sigma\cdot\vec n\,\chi\,
\Phi_0(\epsilon,r,\lambda),
\nonumber\\
\label{320}
 \EA
%
 where
%
 \BE
\lambda=
\frac{2\,a^2(\alpha)}{\epsilon^2}\ .
\label{330}
 \EE
%

The functions $\Phi_{0,1}$ in Eqs.~(\ref{310}) and (\ref{320})
are defined as
%
 \BA
\Phi_0(\epsilon,r,\lambda) =
\frac{1}{4\pi}\int\limits_{0}^{\infty}dt
\!\frac{\lambda}{{\rm sh}(\lambda t)}\,
{\rm exp}\left[- \frac{\lambda\,\epsilon^2 r^2}{4}
{\rm cth}(\lambda t) - t\right]\!,
\nonumber \\
\label{340}
 \EA
%
 \BA
\!\!\vec\Phi_1(\epsilon,r,\lambda)\!=\!
\frac{\epsilon^2\vec r}{8\pi}\!\int\limits_{0}^{\infty}\!\!dt
\!\!\left[\frac{\lambda}{{\rm sh}(\lambda t)}\right]^2
\!\!{\rm exp}\left[- \frac{\lambda\,\epsilon^2 r^2}{4}
{\rm cth}(\lambda t) - t\right]\!,
\nonumber \\
\label{350}
 \EA
%
where $sh(x)$ and $cth(x)$ are the hyperbolic sine and hyperbolic
cotangent, respectively. Note that the $\bar q-q$ interaction enters
Eqs.~(\ref{310}) and (\ref{320}) via the parameter $\lambda$ defined
in (\ref{330}). In the limit of vanishing interaction $\lambda\to 0$
(i.e. $Q^2\to \infty$, $\alpha$ fixed, $\alpha\not=0$ or $1$),
Eqs.~(\ref{310}) and (\ref{320})  produce the perturbative
expressions of Eq.~(\ref{70}). As mentioned above, for charmonium
and bottonium production nonperturbative interaction effects are
weak. Consequently, the parameter $\lambda$ is then rather small due
to the large quark masses $m_c = 1.5$\,GeV and $m_b = 5.0$\,GeV, and
it is given by
%
 \BE
\lambda=
\frac{8\,a^2(\alpha)}{Q^2 + 4~m_{c,b}^{2}}\ .
\label{355}
 \EE
%

With the choice $a^2(\alpha)\propto \alpha(1-\alpha)$ the end-point
behavior of the mean square interquark separation $\la r^2\ra
\propto
1/\alpha(1-\alpha)$ contradicts the idea of confinement. Following
\cite{kst2} we fix this problem via a simple modification of the LC
potential,
%
 \BE
a^2(\alpha) = a^2_0 +4a_1^2\,\alpha(1-\alpha)\ .
\label{180}
 \EE
%
 The parameters $a_0$ and $a_1$ were adjusted in \cite{kst2} to data on
total photoabsorption cross section \cite{gamma1,gamma2}, diffractive
photon dissociation and shadowing in nuclear photoabsorption reactions.
The results of our calculations vary within only $1\%$ when $a_0$ and
$a_1$ satisfy the relation,
%
 \BA
a_0^2&=&v^{ 1.15}\, (0.112)^2\,\GeV^{2}\nonumber\\
a_1^2&=&(1-v)^{1.15}\,(0.165)^2\,\GeV^{2}\ ,
\label{190}
 \EA
%
 where $v$ takes any value $0<v<1$. In
the view of this insensitivity of the observables we fix the
parameters at $v=1/2$. We checked that this choice does not affect
our results beyond a few percent uncertainty.

\subsection{Vector meson wave function}

The last ingredient in the elastic production amplitude (\ref{120})
is the vector meson wave function. We use the popular prescription
of Ref.~ \cite{terentev}, obtained by applying a Lorentz boost to
the rest frame wave function, assumed to be Gaussian, which in turn
leads to radial parts of transversely and longitudinally polarized
mesons in the form
%
 \BE
\Phi_V^{T,L}(\vec r,\alpha) =
C^{T,L}\,\alpha(1-\alpha)\,f(\alpha)\,
{\rm exp}\left[-\ \frac{\alpha(1-\alpha)\,{\vec r}^2}
{2\,R^2}\right]\
\label{170}
 \EE
%
 with the normalization defined below, and
%
 \BE
f(\alpha) = \exp\left[-\ \frac{m_q^2\,R^2}
{2\,\alpha(1-\alpha)}\right]\
\label{170a}
 \EE
%
with the parameters, taken from Ref.~\cite{jan97},
$R=0.515\fm$, $m_q = 0.1\GeV$ for $\rho^0$ production;
$R=0.415\fm$, $m_q = 0.3\GeV$ for $\Phi^0$ production;
$R=0.183\fm$, $m_q = 1.3\GeV$ for charmonium production;
and
$R=0.061\fm$, $m_q = 5.0\GeV$ for bottonium production.

We assume that the distribution amplitude of the $\bar qq$
fluctuations for both the vector meson and the photon have a similar
structure \cite{jan97}.  Then in analogy to Eqs.~(\ref{310}) --
(\ref{320}),
%
 \BE
\Psi^T_V(\vec r,\alpha) =
(E + \vec F\cdot{\vec\nabla}_r)\,
\Phi^T_V(\vec{r},\alpha)\ ,
\label{172}
\EE
%
\BE
\Psi^L_V(\vec r,\alpha) =
2\,m_V\,\alpha(1-\alpha)\,
(\bar\chi\,\vec\sigma\cdot\vec n\,\chi)\,
\Phi^L_V(\vec r,\alpha)\ .
\label{174}
 \EE
%

Correspondingly, the normalization conditions for the transverse and
longitudinal vector meson wave functions read,
%
 \BA
N_{C}\,\int d^{2} r\,\int d\alpha\,
\biggl [\, m_{q}^{2}\,
\Bigl|\Phi^T_{V}(\vec r,\alpha)\Bigr|^{2}
\nonumber \\
+ \Bigl[\alpha^{2} +
(1-\alpha)^{2}\Bigr]\,
\Bigl|\,\partial_{r}
\Phi^T_{V}(\vec r,\alpha)\Bigr|^{2} \biggr ] &=& 1,
\label{230}
 \EA
%
and
%
 \BE
4\,N_{C}\,\int d^{2} r\,\int d\alpha\,
\alpha^{2}\,(1-\alpha)^{2}\,
m_{V}^{2}\,\Bigl|\Phi^L_{V} (\vec r,\alpha)\Bigr|^2 = 1\, .
\label{240}
 \EE
%

%
%
\section{Electroproduction of vector mesons on a nucleon, comparison
with data}\label{data-N}
%
%

As the first test of the formalism, in this section we verify the LC
approach by comparing its results with data for nucleon targets.
Since the expected new HERMES data will be, at separate
polarizations L and T, predominantly for electroproduction of
$\rho^0$, we focus our attention on the production of light vector
mesons. Using all the ingredients specified in the previous
Sect.~\ref{lcc}, i.e. the nonperturbative photon Eqs.~(\ref{310}),
(\ref{320}) and vector meson Eqs.~(\ref{172}), (\ref{174}) wave
functions, we can calculate the forward production amplitude
$\gamma^*\,N \to V\,N$ for transverse and longitudinal photons and
vector mesons. Assuming $s$- channel helicity conservation (SCHC),
the forward scattering amplitude reads,
%
%
\begin{widetext}
 \BA
{\cal M}_{\gamma^{*}N\rightarrow V\,N}^{T}(s,Q^{2})
\Bigr|_{t=0} &=&
N_{C}\,Z_{q}\,\sqrt{2\,\alpha_{em}}
\int d^{2} r\,\sigma_{\bar qq}(\vec r,s)
\int\limits_0^1 d\alpha \Bigl\{ m_{q}^{2}\,
\Phi_{0}(\epsilon,\vec r,\lambda)\Phi^T_{V}(\vec r,\alpha)
\nonumber\\
&-& \bigl [\alpha^{2} + (1-\alpha)^{2}\bigr ]\,
\vec{\Phi}_{1}(\epsilon,\vec r,\lambda)\cdot
\vec{\nabla}_{r}\,\Phi^T_{V}(\vec r,\alpha) \Bigr\}\,;
\label{360}
 \EA
%
 \BA
{\cal M}_{\gamma^{*}N\rightarrow V\,N}^{L}(s,Q^{2})
\Bigr|_{t=0} &=&
4\,N_{C}\,Z_{q}\,\sqrt{2\,\alpha_{em}}\,m_{V}\,Q\,
\int d^{2} r\,\sigma_{\bar qq}(\vec r,s)
\int\limits_0^1 d\alpha\,
\alpha^{2}\,(1-\alpha)^{2}\,
\Phi_{0}(\epsilon,\vec r,\lambda)
\Phi^L_{V}(\vec r,\alpha)\ .
\label{370}
 \EA
\end{widetext}
%
%
 These amplitudes are normalized as ${|{\cal M}^{T,L}|^{2}}=
\left.16\pi\,{d\sigma_{N}^{T,L}/ dt}\right|_{t=0}$, and their real
parts are included according to the prescription described in
Sect.~\ref{lcc}. The terms $\propto \Phi_0(\epsilon,\vec
r,\lambda)\,\Phi_V(\vec r,\alpha)$ and $\propto
\vec{\Phi}_1(\epsilon,\vec r,\lambda)\cdot \vec{\nabla}_r
\Phi_V(\vec r,\alpha)$ in Eqs.~(\ref{360}) and (\ref{370})
correspond to the helicity conserving and helicity-flip transitions
in the $\gamma^*\to \bar qq$, $V\to \bar qq$ vertices, respectively.
The helicity flip transitions represent the relativistic
corrections. For heavy quarkonium these corrections become important
only at large $Q^2\gg m_V^2$. For production of light vector mesons,
however, they are non-negligible even in the photoproduction limit,
$Q^2 = 0$.

Usually the data are presented in the form of the production cross
section $\sigma = \sigma^T + \epsilon\,'\,\sigma^L$, at fixed
photon polarization $\epsilon\,'$. Here the cross section integrated
over $t$ reads:
%
 \beq
\sigma^{T,L}(\gamma^{*}N\to VN) =
\frac{|{\cal M}^{T,L}|^{2}}
{16\pi\,B_{\gamma^*N}}\ ,
\label{375}
 \eeq
%
where $B_{\gamma^*N}\equiv B$ is the slope parameter in the reaction
$\gamma^*\,p \to V\,p$. The absolute value of the production cross
section has been already checked by comparing with data for elastic
$\rho^0$ and $\Phi^0$ electroproduction in Ref.~\cite{knst-01} and
for charmonium exclusive electroproduction $\gamma^*\,p \to
J/\Psi\,p$ in Refs.~\cite{n-02,n-03}.

Motivated by the expected data from the HERMES collaboration, we are
going to make predictions for the production cross sections
$\sigma^{L,T}(\gamma^{*}N\to VN)$ at separate polarizations L and T.
However, the data are usually presented as the ratio $R_{LT} =
\sigma_{L}(\gamma^*N\rightarrow V\,N)/
\sigma_{T}(\gamma^*N\rightarrow V\,N)$ at different photon
virtualities $Q^2$. Then a deviation of $R_{LT}$ from unity
indicates a difference in the production mechanisms of L and T
polarized vector mesons. In order to calculate the ratio $R_{LT}$,
using Eqs.~(\ref{360}) and (\ref{370}) for forward production
amplitudes at different polarizations L and T, one should know
corresponding slope parameters $B_{\gamma_L^*N}\equiv B_L$ and
$B_{\gamma_T^*N}\equiv B_T$:
%
 \beq
R_{LT} =
\frac{|{\cal M}^L|^2}
     {|{\cal M}^T|^2}
\frac{B_T}
     {B_L} \approx
\frac{|{\cal M}^L|^2}
     {|{\cal M}^T|^2}
\biggl (1 + \frac{\Delta B_{TL}}
     {B}\biggr )\, ,
\label{377}
 \eeq
%
where $\Delta B_{TL} = B_T - B_L$.
%
  \begin{figure}[bht]
\includegraphics{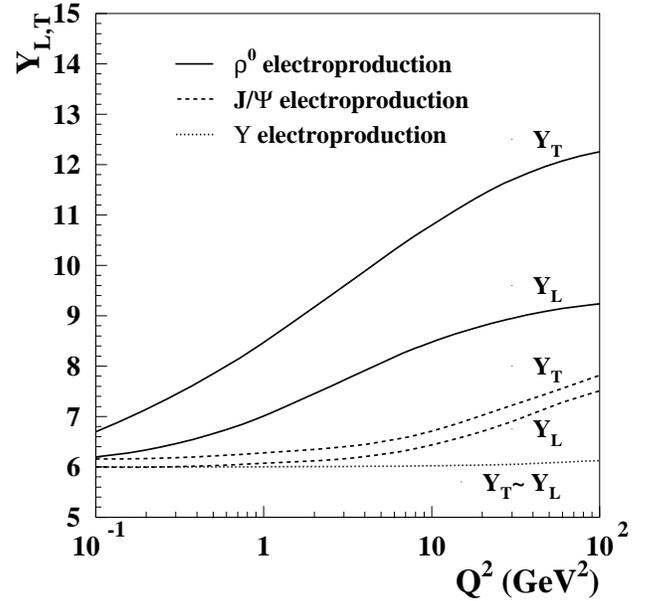}
\begin{center}
\vspace{8.0cm}
{\caption[Delta]
 {\sl $Q^2$ dependence of the scale parameters $Y_{L}$
and $Y_{T}$, from the expression for the scanning radius (\ref{10}),
corresponding to the production of $L$ and $T$ polarized vector
mesons. Solid, dashed and dotted lines represent electroproduction
of $\rho^0$, $J/\Psi$ and $\Upsilon$, respectively. }
 \label{y-lt}}
 \end{center}
 \end{figure}
%

The scanning phenomenon, Eq.~(\ref{10}), was already discussed in
Refs.~\cite{bfkl94,jan97,nnpzz-98}, and it can be understood
qualitatively by analyzing the forward production amplitude
(\ref{120}). Here we assume for simplicity the perturbative
distribution amplitudes of the $\bar qq$ Fock component of the
photon containing the Bessel function $K_0(\epsilon r)$ (see
Eq.~(\ref{70})). As was mentioned in the previous Sect.~\ref{lcc},
the most important property of the dipole cross section
$\sigma_{\bar qq}(\vec r,s)$ is the color transparency driven
dependence $\propto r^2$ at small $r$. Due to the smooth shape of
the vector meson wave functions $\Phi_V^{L,T}(r,\alpha)$ (see
Eq.~(\ref{170})) and because of the behavior of the Bessel functions
$K_{0,1}(x)\propto exp(-x)$ at large values of $x$, the production
amplitude is dominated by the contribution from $r_S\approx
3/\epsilon$. In the nonrelativistic approximation of $m_V\sim
2\,m_q$ and $\alpha\sim 0.5$, it leads to the scanning radius
(\ref{10}) and the estimate $Y\approx 6$ (see also Fig.~\ref{y-lt}).

In general, to be more precise, the scanning property (see
Eq.~(\ref{10}) is quantified separately for L and T polarizations
via the $Q^2$- dependent scale parameters $Y_L$ and $Y_T$, as
illustrated in Fig.~\ref{y-lt}. The dotted line represents the fact
that for electroproduction of bottonia both scale parameters
$Y_T\sim Y_L\sim 6$, and practically do not depend on $Q^2$ as a
consequence of the nonrelativistic approximation. Dashed lines
describe the $Q^2$ dependence of $Y_{L}$ and $Y_{T}$ for charmonium
electroproduction. One can see that both $Y_L$ and $Y_T$ smoothly
rise with $Q^2$, do not differ much to each other and are a little
bit higher than the value $6$ resulting from the nonrelativistic
approximation. For this reason charmonium can be safely treated as a
nonrelativistic object at small and medium values of $Q^2$ such that
$r_S \gsim R_{J/\Psi}$\footnote{ $R_{J/\Psi}$ is the radius of
charmonium}. At larger $Q^2\gg m_{J/\Psi}^2$, the scale parameters
$Y_{L,T}$ have a stronger $Q^2$ dependence reaching the value $\sim
7.7$ at $Q^2 = 100\,GeV^2$, which differs from the nonrelativistic
value $Y\sim 6$. It means that relativistic effects are no longer
negligible and should be included in the calculations
\cite{nnpzz-98}. However, the situation is completely different for
light vector meson production, as illustrated in Fig.~\ref{y-lt} by
the solid lines. In this case, due to the presence of strong
relativistic effects, the scale parameters rise with $Q^2$ much more
rapidly than for the production of heavy vector mesons.

Compared to ${\cal M}_{L}$ (\ref{360}) the transverse production
amplitude ${\cal M}_{T}$ (\ref{370}) receives larger contributions
from large-size asymmetric end-point $\bar qq$ fluctuations with
$\alpha\,(1 - \alpha)\ll 1$. This is illustrated in Fig.~\ref{y-lt},
where $Y_T > Y_L$ in the whole $Q^2$-range, and the difference
between $Y_T$ and $Y_L$ rises with $Q^2$. This fact is especially
evident for the electroproduction of $\rho^0$ mesons, depicted by
the solid lines. For electroproduction of charmonia the difference
between $Y_{T}$ and $Y_{L}$ is small as a consequence of small
relativistic effects, while for electroproduction of bottonia it was
already indicated in Ref.~\cite{nnpzz-98} that $Y_T\sim Y_L\sim 6$
in a very broad $Q^2$-range, which supports the conclusion that the
relativistic corrections are negligible.

At small $r_S \lsim R_V$, the production
amplitudes (\ref{360}) and (\ref{370}) can be evaluated as
%
 \beq
{\cal M}_{T} \propto r_S^2\,\sigma_{\bar qq}(r_S,s) \propto
\frac{Y_T^4}{(Q^2 + m_V^2)^2}\, ,
\label{378}
 \eeq
%
%
 \beqn
{\cal M}_{L} \propto \frac{\sqrt{Q^2}}{m_V}\,r_S^2\,\sigma_{\bar
qq}(r_S,s) & \propto &
\frac{\sqrt{Q^2}}{m_V}\,\frac{Y_L^4}{(Q^2 + m_V^2)^2}
\nonumber \\
\propto
\frac{\sqrt{Q^2}}{m_V}\,\frac{Y_L^4}{Y_T^4}\,{\cal M}_{T}\, ,
\label{379}
 \eeqn
%
which means that the longitudinally polarized vector mesons
dominate at $Q^2\gg m_V^2$.

A detailed analysis of the diffraction cone \cite{nnpzz-98,n-01} for
exclusive vector meson electroproduction, within the color dipole
generalized Balitskij-Fadin-Kuraev-Lipatov (BFKL) phenomenology,
showed the presence of geometrical contributions from the target
nucleon $\sim B_N$ and the beam dipole $\sim r^2$. At fixed energy
and according to the scanning phenomenon (\ref{10}), the diffraction
slope is predicted to decrease with $(Q^2 + m_V^2)$ as
%
\BE B(Q^2)\sim B_N + \tilde{C}\,r_S^2 \approx B_N +
const\,\frac{Y^2}{Q^2 + m_V^2}\, . \label{377B} \EE
%

One can see from Eq.~(\ref{377B}) that different scanning properties
for L and T polarized vector mesons ($Y_L < Y_T$, see
Fig.~\ref{y-lt} and subsequent discussion) lead also to an
inequality $B_L < B_T$ of the slope parameters in the reactions
$\gamma_L^{*}N\to V_L\,N$ and $\gamma_T^{*}N\to V_T\,N$.
Consequently, the difference $\Delta B_{TL}$ in Eq.~(\ref{377}) is
positive and can be estimated as
%
\BE
\Delta B_{TL}\propto
\frac{\Delta Y_{TL}^2}{Q^2 + m_V^2}\, ,
\label{377BB}
\EE
%
where
%
\BE
\Delta Y_{TL}^2 = Y_T^2 - Y_L^2\, .
\label{377BBB}
\EE
%

For electroproduction of $\rho^0$ mesons at small $Q^2\lsim
m_{\rho}^2$ the rise of $\Delta Y_{TL}^2$ with $Q^2$ can compensate
or even overcompensate the decrease of $\Delta B_{TL}$ with $(Q^2 +
m_{\rho}^2)$. Consequently, the difference $\Delta B_{TL}$ in
Eq.~(\ref{377BB}) can weakly rise with $Q^2$. This does not happen
at larger $Q^2 > m_{\rho}^2$, when $\Delta B_{TL}$ decreases slowly
with $Q^2$. In the HERMES kinematical range $\Delta B_{TL}\sim
0.7\GeV^{-2}$ at $Q^2 = 0.7\GeV^2$, reaching the value $\sim
0.4\GeV^{-2}$ at $Q^2 = 5\GeV^2$. Correspondingly, the factor
$\Delta B_{TL}/B$ in Eq.~(\ref{377}), treated as a correction to
unity in the brackets, is about $0.09$ at $Q^2 = 0.7\GeV^2$ and
decreases very slowly with $Q^2$ reaching a value of $\sim 0.07$ at
$Q^2 = 5\GeV^2$. For this reason the factor $\Delta B_{TL}/B$ cannot
be neglected in the calculations.

Using Eqs.~(\ref{377}) and (\ref{379}) one can present the nucleon
$L/T$-ratio as
%
 \beq
R_{LT} \propto
\frac{Q^2}{m_V^2}\,
\frac{Y_L^8}{Y_T^8}\,
\frac{B_T}
     {B_L} \approx
\frac{Q^2}{m_V^2}\,
F_Y(Q^2)\,
\biggl (1 + \frac{\Delta B_{TL}}
     {B}\biggr )\, ,
\label{544}
 \eeq
%
and thus $R_{LT}$ is given mainly by three ingredients.

i.) The presence of the factor $Q^2/m_V^2$, which comes from
$\sigma_L$ (see Eq.~(\ref{370})), and represents a generic
consequence of electromagnetic gauge invariance.

ii.) The presence of the $Q^2$-dependent factor $F_Y(Q^2) =
Y_L^8/Y_T^8$, which comes from the scanning phenomenon (\ref{10}),
and reflects the different relativistic corrections for the L and T
production amplitudes. These corrections become important only at
large $Q^2\gg m_V^2$ (see Fig.~\ref{y-lt}). The factor $F_Y$ leads
to a substantial reduction of the rise of $R_{LT}$ with $Q^2$,
especially for the production of light vector mesons.

iii.) The presence of the factor $B_T/B_L$, which follows from the
fact that the slope parameters $B_L$ and $B_T$ for the production of
L and T polarized vector mesons are different. According to the
scanning property ($B_L < B_T$), this factor  decreases slightly
with $Q^2$ tending to unity from above at large $Q^2\gg m_V^2$
\cite{nnpzz-98}. The ratio $B_T/B_L$ leads to an additional but
small reduction of the $Q^2$- rise of $R_{LT}$.
%
  \begin{figure}[bht]
\includegraphics{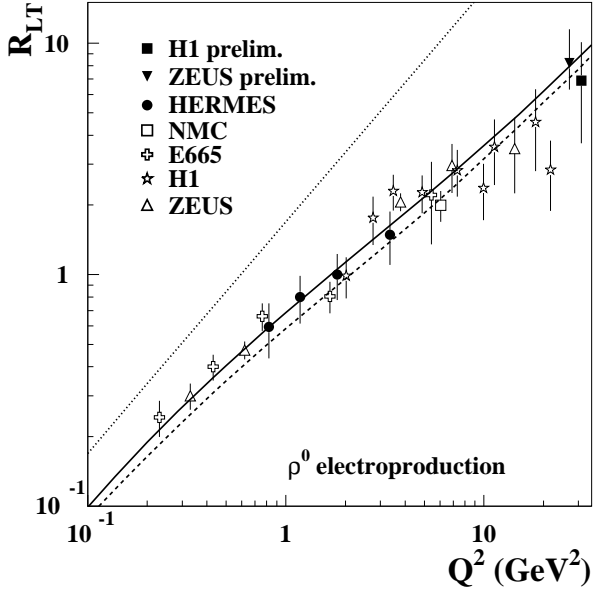}
\begin{center}
\vspace{8.0cm}
{\caption[Delta]
 {\sl $Q^2$ dependence of the ratio $R_{LT}$
of the integrated cross sections for the reactions $\gamma_L^*\,p
\to \rho^0_L\,p$ and $\gamma_T^*\,p \to \rho^0_T\,p$. The solid and
dashed lines represent model calculations at $W = 15$ and $90\GeV$,
respectively. The data are taken from \cite{hermes-lts}. The
preliminary H1 and ZEUS data can be found in \cite{h1-ro-prel} and
\cite{zeus-ro-prel}, respectively. The dotted curve represents the
$Q^2/m_V^2$- rise of $R_{LT}$. }
 \label{r-lt-q2-n}}
 \end{center}
 \end{figure}
%

%
  \begin{figure}[bht]
\includegraphics{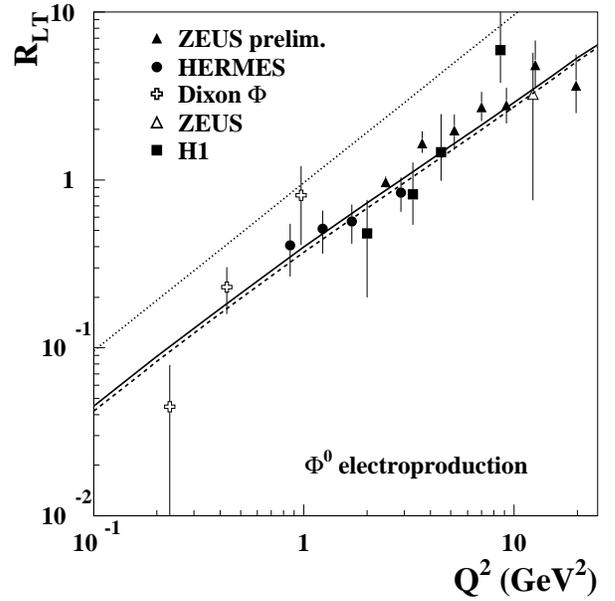}
\begin{center}
\vspace{8.0cm}
{\caption[Delta]
 {\sl $Q^2$ dependence of the ratio $R_{LT}$
of the integrated cross sections for the reactions $\gamma_L^*\,p
\to \Phi^0_L\,p$ and $\gamma_T^*\,p \to \Phi^0_T\,p$. The solid and
dashed lines represent model calculations at $W = 15$ and $90\GeV$,
respectively. The data are taken from \cite{hermes-lts}. and H1 data
from \cite{h1-lt1}. The preliminary ZEUS data can be found in
\cite{zeus-fi-prel}. The dotted curve represents the $Q^2/m_V^2$-
rise of $R_{LT}$. }
 \label{r-lt-q2-n1}}
 \end{center}
 \end{figure}
%

\begin{figure}[bht]
\includegraphics{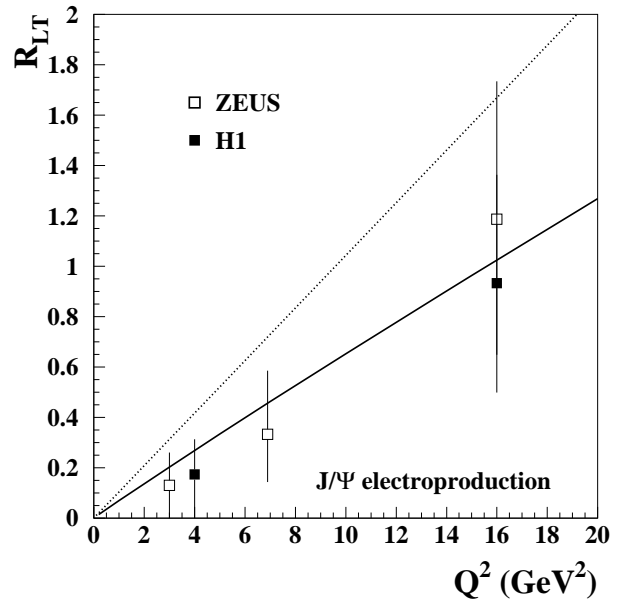}
\begin{center}
\vspace{8.0cm}
{\caption[Delta]
 {\sl $Q^2$ dependence of the ratio $R_{LT}$
of the integrated cross sections for
the reactions $\gamma_L^*\,p \to {(J/\Psi)}_L\,p$
and $\gamma_T^*\,p \to {(J/\Psi)}_T\,p$.
The model calculations are performed
at $W = 90\,\GeV$.
The H1 and ZEUS data are taken
from \cite{h1-jpsi-q2} and \cite{zeus-jpsi-q2},
respectively.
The dotted curve represents
the $Q^2/m_V^2$- rise of $R_{LT}$.
}
 \label{r-lt-q2-n2}}
 \end{center}
 \end{figure}

Our predictions are plotted in Figs.~\ref{r-lt-q2-n} and
\ref{r-lt-q2-n1}, together with the data on the $Q^2$ dependence of
the ratio $R_{LT}$ for the production of $\rho^0$ and $\Phi^0$
mesons, taken from Ref.~\cite{hermes-lts}. We added also the last
published data from H1 \cite{h1-lt1} and preliminary data from the
H1 \cite{h1-ro-prel} and ZEUS\cite{zeus-ro-prel,zeus-fi-prel}
collaborations. The analogous $Q^2$ dependence of $R_{LT}$ for
electroproduction of charmonia is plotted in Fig.~\ref{r-lt-q2-n2}
together with results from the H1 \cite{h1-jpsi-q2} and ZEUS
\cite{zeus-jpsi-q2} collaborations.
%
%

One can see from Eq.~(\ref{544}) that the $Q^2$-rise of the nucleon
ratio $R_{LT}$ gets diminished by the factor $F_Y$, coming from the
different scanning properties of the L and T production amplitudes,
and by the ratio of the slope parameters $B_T$ and $B_L$. In the
nonrelativistic approximation, represented by the electroproduction
of bottonia, the factor $F_Y\sim 1$ (see Eq.~(\ref{544}) and
Fig.~\ref{y-lt}) and $B_T\sim B_L$. Consequently, the ratio $R_{LT}$
rises with $Q^2$ as $\sim Q^2/m_V^2$ \cite{nnpzz-98}. On the other
hand, electroproduction of light vector mesons has large
relativistic effects and the different scanning properties for the L
and T production amplitudes ($Y_L < Y_T$) lead to a large decrease
of the dominance of the longitudinal cross section $\propto |{\cal
M}_L|^2$. Thus the ratio $R_{LT}$ rises with $Q^2$ much less rapidly
than $Q^2/m_V^2$. This is illustrated in Figs.~\ref{r-lt-q2-n} and
\ref{r-lt-q2-n1} as a difference between the solid (dashed) and
dotted lines. For charmonium production the decrease of the rise of
the ratio $R_{LT}$ with $Q^2/m_{J/\Psi}^2$ is much less effective
due to smaller relativistic effects, as one can see in
Fig.~\ref{r-lt-q2-n2} as a difference between the solid and dotted
lines.

Notice that in all calculations we assumed SCHC as the consequence
of the spin independence of the dipole cross section $\sigma_{\bar
qq}(r,s)$ in the forward production amplitude (\ref{120}). This
assumption is supported by the low-energy data, indicating that the
amplitude for the photon/vector-meson transition is predominantly
$s$- channel conserving, i.e. the helicity of the vector meson is
equal to that of the photon when the spin-quantization axis is
chosen along the direction of the meson momentum in the
$\gamma^*\,p$ center-of-mass system. In general, however, small
helicity-single-flip and helicity-double-flip contributions to the
production amplitude have been reported in $\pi^+\,\pi^-$
photoproduction in the $\rho^0$ mass region, at $W\lsim 4\GeV$
\cite{sfa-sbt}. Helicity-single-flip amplitudes have also been
observed in $\rho^0$ electroproduction for $1.3 < W < 2.8\GeV$ and
$0.3 < Q^2 < 1.4\GeV^2$ \cite{sfa-joos}. A helicity-single-flip
contribution of $(14\,\pm 8)\,\%$ was measured in $\rho^0$
muo-production at $W = 17\GeV$ \cite{chio}.

At high energy the breaking of SCHC has been measured by the ZEUS
\cite{sfa-zeus} and H1 \cite{h1-lt2} collaborations at HERA. The
size of the SCHC-breaking effects was quantified by evaluating the
ratios of the helicity-single-flip and helicity-double-flip
amplitudes to the helicity-conserving amplitudes. The ratio of
$T_{01}$ (for production of L polarized $\rho^0$ mesons from T
photons) to the helicity-conserving amplitudes,
%
 \beq
\tau_{01} =
\frac{|T_{01}|}{\sqrt{|T_{00}|^2 + |T_{11}|^2}}\, ,
\label{380}
 \eeq
%
gives the values $\tau_{01} = (6.9\pm 2.7)\,\%$ for $0.25 < Q^2 <
0.85\GeV^2$ and $\tau_{01} = (7.9\pm 2.6)\,\%$ for $3 < Q^2 <
30\GeV^2$, determined by the ZEUS collaboration \cite{sfa-zeus}. The
H1 result for this quantity is $(8\,\pm 3)\,\%$ \cite{h1-lt2}. The
ratio of helicity-double-flip amplitudes to the helicity-conserving
amplitudes $\tau_{1\,-1}$, defined analogously as $\tau_{01}$ in
Eq.~(\ref{380}) gives the values $\tau_{1\,-1} = (4.8\pm 2.8)\,\%$
for $0.25 < Q^2 < 0.85\GeV^2$ and $\tau_{1\,-1} = (1.4\pm 6.5)\,\%$
for $3 < Q^2 < 30\GeV^2$ \cite{sfa-zeus}. Besides, the ZEUS
collaboration \cite{sfa-zeus} also determined the nucleon $L/T$-
ratio for $\rho^0$ electroproduction, without assuming SCHC. The
results thus found differ from those derived from the SCHC
hypothesis by less than 3$\,\%$. Similarly the last data from the
HERMES collaboration \cite{hermes-lts} on electroproduction of
$\rho^0$ and $\Phi^0$ mesons at $4 < W < 6\GeV$ and $0.7 < Q^2 <
5\GeV^2$ confirm the breaking of SCHC and are consistent with the H1
\cite{h1-lt2} and ZEUS \cite{sfa-zeus} results. The observed
deviation from SCHC changes the ratio $R_{LT}$ by only a few
percent. Because in the present paper we will focus predominantly on
theoretical predictions for the ratio of the L and T production
cross sections on nucleon and  nuclear targets we can safely assume
SCHC.

The second test of our approach is the description of the energy
dependence of the production amplitudes (\ref{360}) and (\ref{370}),
which is given by the energy dependent dipole cross section. As we
mentioned in the previous section $\sigma_{\bar qq}(r,s)$ has a
stronger energy dependence at smaller dipole sizes. According to the
scanning phenomenon discussed above, the dipole cross section is
scanned at smaller transverse size in the L than in the T production
amplitude. Consequently, the L production amplitude has a stronger
energy dependence and so we expect a weak energy dependence of the
ratio $R_{LT}$. Model predictions at $W = 15$ and $90\GeV$ are
depicted in Figs.~\ref{r-lt-q2-n} and \ref{r-lt-q2-n1} by the dashed
and solid lines. One can see that the error bars of data are too
large to see such a weak energy dependence.

%
%
\section{Incoherent production of vector mesons off
nuclei}\label{vm-incoh}
%
%

%
\subsection{Introduction}\label{intro-inc}
%

In diffractive incoherent (quasielastic) production of vector mesons
off nuclei, $\gamma^{*}\,A\rightarrow V\,X$, one sums over all final
states of the target nucleus, except those which contain particle
(pion) creation. The observable that is usually studied
experimentally is the nuclear transparency, defined as
%
 \BE
Tr^{inc}_{A} =
\frac{\sigma_{\gamma^{*}A\to VX}^{inc}}
{A\,\sigma_{\gamma^{*}N\to VN}}\ .
\label{480}
 \EE
%
 The $t$-slope of the differential quasielastic cross section is the same
as on a nucleon target. Therefore, instead of integrated cross
sections one can also use the forward differential cross sections
given in Eq.~(\ref{125}), to write,
%
 \beq
Tr^{inc}_A = \frac{1}{A}\,
\left|\frac{{\cal M}_{\gamma^{*}A\to VX}(s,Q^{2})}
{{\cal M}_{\gamma^{*}N\to VN}(s,Q^{2})}\right|^2\, .
\label{485}
 \eeq
%
We consider also the production of either longitudinal or transverse
polarized vector mesons on nucleon and nuclear targets, and then one
can define nuclear transparency separately for incoherent production
of L and T vector mesons as
%
 \beq
Tr^{inc}_A(L) = \frac{1}{A}\,
\left|\frac{{\cal M}_{\gamma_L^{*}A\to V_L\,X}(s,Q^{2})}
{{\cal M}_{\gamma_L^{*}N\to V_L\,N}(s,Q^{2})}\right|^2\,
\label{487}
 \eeq
%
and
%
 \beq
Tr^{inc}_A(T) = \frac{1}{A}\,
\left|\frac{{\cal M}_{\gamma_T^{*}A\to V_T\,X}(s,Q^{2})}
{{\cal M}_{\gamma_T^{*}N\to V_T\,N}(s,Q^{2})}\right|^2\, .
\label{489}
 \eeq
%

However, when one wants to study the ratio of L and T polarized
vector meson production on nuclear targets using forward
differential cross sections we should include also the difference
between the L and T slope parameters, as was done in the previous
Section (see Eq.~(\ref{377})) :
%
 \BA
R_{LT}^{A}(inc) &=&
\frac{\sigma_{\gamma_L^{*}A\to V_L\,X}^{inc}}
{\sigma_{\gamma_T^{*}A\to V_T\,X}^{inc}} =
\left|\frac{{\cal M}_{\gamma_L^{*}A\to V_L\,X}(s,Q^{2})}
{{\cal M}_{\gamma_T^{*}A\to V_T\,X}(s,Q^{2})}\right|^2
\frac{B_T}{B_L}
\nonumber \\
&=&
R_{LT}\,\frac{Tr_A^{inc}(L)}{Tr_A^{inc}(T)}
= R_{LT}\,f_{inc}(s,Q^2,A)\, ,
\label{490}
 \EA
%
where the nuclear transparencies $Tr_A^{inc}(L)$ and $Tr_A^{inc}(T)$
for L and T polarized vector mesons are given by Eqs.~(\ref{487})
and (\ref{489}), respectively. The variable $f_{inc}$ in
Eq.~(\ref{490}) represents the nuclear modification factor already
introduced by Eq.~(\ref{65}).

%
\subsection{The LC Green function formalism}\label{lc-inc}
%

The nuclear forward production amplitude ${\cal M}_{\gamma^{*}\,A\to
V\,X}(s,Q^{2})$ was calculated using the LC Green function approach
in Ref.~\cite{knst-01}. In this approach the physical photon
$|\gamma^*\ra$ is decomposed into different Fock states, namely, the
bare photon $|\gamma^*\ra_0$, plus $|\bar qq\ra$, $|\bar qqG\ra$,
etc. As we mentioned above the higher Fock states containing gluons
describe the energy dependence of the photoproduction reaction on a
nucleon. Besides, these Fock components also lead to gluon shadowing
as far as nuclear effects are concerned. However, these fluctuations
are heavier and have a shorter coherence time (lifetime) than the
lowest $|\bar qq\ra$ state, and therefore at medium energies only
the $|\bar qq\ra$ fluctuations of the photon matter. Consequently,
gluon shadowing, related to the higher Fock states, will dominate at
high energies. A detailed description and calculation of gluon
shadowing for the case of vector meson production off nuclei is
presented in \cite{knst-01,ikth-02}. In the HERMES kinematical range
studied in the present paper the gluon shadowing is negligible and
therefore is not included in the calculations.

The propagation of an interacting $\bar qq$ pair in a nuclear medium
is described by the Green function satisfying the evolution
Eq.~(\ref{250}). However, the potential in this case acquires an
imaginary part which represents absorption in the medium (see
Eq.~(\ref{40}) for notation),
%
 \BE
Im V_{\bar qq}(z_2,\vec r,\alpha) = -
\frac{\sigma_{\bar qq}(\vec r,s)}{2}\,\rho_{A}({b},z_2)\,.
\label{440}
 \EE
%
 The evolution equation (\ref{250}), with the potential
$V_{\bar qq}(z_{2},\vec r_{2},\alpha)$ containing this imaginary
part, was used in \cite{krt1,krt2}. In particular nuclear shadowing
in deep-inelastic scattering was calculated, in good agreement with
data.

Analytical solutions of Eq.~(\ref{250}) are only known for the
harmonic oscillator potential $V(r)\propto r^2$. Furthermore, to
keep the calculations reasonably simple we use the dipole
approximation,
%
 \beq
\sigma_{\bar qq}(r,s) = C(s)\,r^2\ ,
\label{460}
 \eeq
%
which allows to obtain the Green function in an analytical form
(see Eq.~(\ref{270})).

The energy dependent factor $C(s)$ was adjusted by demanding that
the calculations employing the approximation Eq.~(\ref{460})
reproduce correctly the results based on the realistic cross section
(\ref{130}), in the limit $l_c\gg R_A$ (the so called ``frozen''
approximation), when the Green function takes the simple form :
%
 \BA
&&G_{\bar qq}(z_1,\vec r_1;z_2,\vec r_2) \Rightarrow
\nonumber \\
&&\delta(\vec r_1-\vec r_2)\,\exp\left[
-{1\over2}\,\sigma_{\bar qq}(r_1)
\int\limits_{z_1}^{z_2} dz\,\rho_A(b,z)\right]\ ,
\label{465}
 \EA
%
where the dependence of the Green function on impact parameter has
been dropped. A detailed description of the determination of the
factors $C(s)$, separately for coherent and incoherent vector meson
production, is presented in the paper \cite{knst-01}.

With the potential given by the Eqs.~(\ref{440}) -- (\ref{460}), the
solution of Eq.~(\ref{250}) has the same form as Eq.~(\ref{270}),
except that one should replace $\omega \Rightarrow \Omega$, where
%
 \beq
\Omega = \frac{\sqrt{a^4(\alpha)-
i\,\rho_{A}({b},z)\,
\nu\,\alpha\,(1-\alpha)\,C(s)}}
{\nu\;\alpha(1-\alpha)}\ .
\label{470}
 \eeq
%

The evolution equation (\ref{250}), with the potential $V_{\bar
qq}(z_{2},\vec r_{2},\alpha)$ containing the imaginary part
(\ref{440}), and with the realistic dipole cross section
(\ref{130}), was recently solved numerically for the first time in
Ref.~\cite{dis-num}. There it was shown that the nuclear shadowing
in deep-inelastic scattering depends on the form of the dipole cross
section $\sigma_{\bar qq}$. However, the approximation (\ref{460})
gives a nuclear shadowing which is very close to realistic numerical
calculations using the parametrization (\ref{130}), in the HERMES
kinematical range under consideration in the present paper. For this
reason we can safely use the dipole approximation (\ref{460}) for
the calculation of vector meson production.

%
\subsection{Different regimes for incoherent production of vector
mesons}\label{regimes}
%

As we discussed in \cite{knst-01}, the value of $l_c$ can
distinguish different regimes of vector meson production.

{\bf (i)} The CL is much shorter than the mean nucleon spacing in a
nucleus ($l_c \to 0$). In this case $G(z_2,\vec r_2;z_1,\vec r_1)
\to \delta(z_2-z_1)$. Correspondingly, the formation time of the
meson wave function is very short, and it is given by
Eq.~(\ref{20}). For light vector mesons $l_f\sim l_c$, and since the
formation and coherence lengths are proportional to the photon
energy, both must be short. Consequently, nuclear transparency is
given by the simple formula Eq.~(\ref{40}) corresponding to the
Glauber approximation.

{\bf (ii)} In the production of charmonia and other heavy flavors,
the intermediate case $l_c\to 0$, but $l_f\sim R_A$, can be
realized. Then the formation of the meson wave function is described
by the Green function, and the numerator of the nuclear transparency
ratio, Eq.~(\ref{485}), has the form \cite{kz-91},
%
 \BA
\Bigl|{\cal M}_{\gamma^{*}A\to VX}(s,Q^{2})
\Bigr|^2_{l_c\to0;\,l_f\sim R_A} =
\nonumber\\
\int d^2b\int_{-\infty}^{\infty} dz\,\rho_A(b,z)\,
\Bigl|F_1(b,z)\Bigr|^2\ ,
\label{500}
 \EA
%
 where
%
 \BA
F_1(b,z) =
\int_0^1 d\alpha
\int d^{2} r_{1}\,d^{2} r_{2}\,
\Psi^{*}_{V}(\vec r_{2},\alpha)\,\times
\nonumber \\
G(z^\prime,\vec r_{2};z,\vec r_{1})\,
\sigma_{\bar qq}(r_{1},s)\,
\Psi_{\gamma^{*}}(\vec r_{1},
\alpha)\Bigl|_{z^\prime\to\infty}
\label{505}
 \EA
%

{\bf (iii)} The high energy limit $l_c \gg R_A$ (in fact, it is more
correct to compare with the mean free path of the $\bar qq$ in a
nuclear medium if the latter is shorter than the nuclear radius). In
this case $G(z_2,\vec r_2;z_1,\vec r_1) \to \delta(\vec r_2 - \vec
r_1)$, i.e. all fluctuations of the transverse $\bar qq$ separation
are ``frozen'' by Lorentz time dilation. Then, the numerator on the
right-hand-side (r.h.s.) of Eq.~(\ref{485}) takes the form
\cite{kz-91},
%
 \beqn
&&\Bigl|{\cal M}_{\gamma^{*}A\to VX}(s,Q^{2})
\Bigr|^2_{l_c \gg R_A}=
\int d^2b\,T_A(b)
\nonumber\\&\times&
\left|\int d^2r\int_0^1 d\alpha
\,\Psi_{V}^{*}(\vec r,\alpha)\,
\sigma_{\bar qq}(r,s)\right.
\nonumber\\&\times& \left.
\exp\left[-{1\over2}\sigma_{\bar qq}(r,s)\,T_A(b)\right]
\Psi_{\gamma^{*}}(\vec r,\alpha,Q^2)\right|^2
\label{510}
 \eeqn
%
In this case the $\bar qq$ pair attenuates with a constant
absorption cross section, like in the Glauber model, except that the
whole exponential is averaged rather than just the cross section in
the exponent. The difference between the results of the two
prescriptions are the well known inelastic corrections of Gribov
\cite{zkl}.

{\bf (iv)} The general case when there are no restrictions for
either $l_c$ or $l_f$. The corresponding theoretical tool has been
developed for the first time only recently in \cite{knst-01}, and
applied to electroproduction of light vector mesons at medium and
high energies. The same approach was used later for the study of
virtual photoproduction of heavy vector mesons \cite{n-02,n-03}.
Even within the VDM the Glauber model expression interpolating
between the limiting cases of low [(i), (ii)] and high [(iii)]
energies has been also derived only recently \cite{hkn}. In this
general case the incoherent nuclear production amplitude squared is
represented as a sum of two terms \cite{hkz},
%
 \BA
\Bigl|\,{\cal M}_{\gamma^{*}A\to
VX}(s,Q^{2})\Bigr|^{2} = \int d^{2}b
\int\limits_{-\infty}^{\infty} dz\,\rho_{A}({b},z)\,
&&\times
\nonumber \\
\Bigl|F_{1}({b},z) - F_{2}({b},z)\Bigr|^{2}\ .
\label{520}
 \EA
%

 The first term $F_{1}({b},z)$, introduced above in Eq.~(\ref{505})
corresponds to the short $l_c$ limit (ii). The second term
$F_{2}({b},z)$ in (\ref{520}) corresponds to the situation when the
incident photon produces a $\bar qq$ pair diffractively and
coherently at the point $z_1$, prior to incoherent quasielastic
scattering at point $z$. The LC Green functions describe the
evolution of the $\bar qq$ over the distance from $z_1$ to $z$ and
further on, up to the formation of the meson wave function.
Correspondingly, this term has the form,
%
 \BA
F_{2}(b,z) &=& \frac{1}{2}\,
\int\limits_{-\infty}^{z} dz_{1}\,\rho_{A}(b,z_1)\,
\int\limits_0^1 d\alpha\int d^2 r_1\,
d^2 r_{2}\,d^2 r\,
\times
\nonumber \\
&&\Psi^*_V (\vec r_2,\alpha)\,
G(z^{\prime}\to\infty,\vec r_2;z,\vec r)\,
\sigma_{\bar qq}(\vec r,s)\,
\times
\nonumber \\
&&G(z,\vec r;z_1,\vec r_1)\,
\sigma_{\bar qq}(\vec r_1,s)\,
\Psi_{\gamma^{*}}(\vec r_1,\alpha)\, .
\label{530}
 \EA
%

Eq.~(\ref{520}) correctly reproduces the limits (i) - (iii). At
$l_c\to 0$ the second term $F_2(b,z)$ vanishes because of strong
oscillations, and Eq.~(\ref{520}) reproduces the Glauber expression
Eq.~(\ref{40}). At $l_c\gg R_A$ the phase shift in the Green
functions can be neglected and they acquire the simple form
$G(z_2,\vec r_2;z_1,\vec r_1) \to \delta(\vec r_2 - \vec r_1)$. In
this case the integration over longitudinal coordinates in
Eqs.~(\ref{505}) and (\ref{530}) can be performed explicitly, and
the asymptotic expression Eq.~(\ref{510}) is recovered as well.
\\

%
\subsection{The nuclear ratio $R_{LT}^{A}(inc)$
in the limit of long coherence length $l_c\gg R_A$}\label{rlt-inc}
%

One can see from Eqs.~(\ref{65}) and (\ref{490}) that the nuclear
ratio $R_{LT}^{A}(inc)$ differs from the nucleon ratio $R_{LT}$ by
the nuclear modification factor for incoherent process
$f_{inc}(s,Q^2,A)$, given also as the ratio
$Tr_A^{inc}(L)/Tr_A^{inc}(T)$ of nuclear transparencies for the
corresponding polarizations L and T. In order to understand more
intuitively and simply the $Q^2$ and $A$ dependence of the nuclear
ratio $R_{LT}^{A}(inc)$, it is convenient to present the nuclear
transparency in the high energy limit $l_c\gg R_A$ (see
Eq.~(\ref{510})):
%
\begin{widetext}
 \beqn
Tr_A^{inc}\Bigr |_{l_c\gg R_A} &=&
\frac{
\int d^2b\,T_A(b)\left|\int d^2r\int_0^1 d\alpha
\,\Psi_{V}^{*}(\vec r,\alpha)\,
\sigma_{\bar qq}(r,s)\,
\exp\left[-{1\over2}\sigma_{\bar qq}(r,s)\,T_A(b)\right]
\Psi_{\gamma^{*}}(\vec r,\alpha,Q^2)\right|^2}
{A\,
\left|\int d^2r\int_0^1 d\alpha
\,\Psi_{V}^{*}(\vec r,\alpha)\,
\sigma_{\bar qq}(r,s)\,
\Psi_{\gamma^{*}}(\vec r,\alpha,Q^2)\right|^2}
\nonumber \\
&=&
1 - \Sigma\,\frac{1}{A}\,\int\, d^2b\,{T_A(b)}^2 + \ldots
\, ,
\label{436}
 \eeqn
\end{widetext}
%
where the CT observable \cite{knnz94}
%
 \beqn
\Sigma =
\frac{
\int d^2r\int_0^1 d\alpha
\,\Psi_{V}^{*}(\vec r,\alpha)\,
\sigma_{\bar qq}^2(r,s)\,
\Psi_{\gamma^{*}}(\vec r,\alpha,Q^2)}
{
\int d^2r\int_0^1 d\alpha
\,\Psi_{V}^{*}(\vec r,\alpha)\,
\sigma_{\bar qq}(r,s)\,
\Psi_{\gamma^{*}}(\vec r,\alpha,Q^2)}
\label{437}
 \eeqn
%
measures the strength of the intranuclear final state interaction
(FSI).

For the sake of clarity in the subsequent discussion we have
explicitly shown in Eq.~(\ref{436}) only the leading term of the
FSI. Evaluation of the strength of the FSI (\ref{437}) can be done
using the scanning phenomenon (\ref{10}) (see also Eq.~(\ref{212})
and Fig.~\ref{y-lt}). Since the integrand of the matrix element in
the numerator of Eq.~(\ref{437}) is peaked at $r\sim r_{FSI} = 5/3\,
r_S$\footnote{Extension to the higher-order rescattering is
straightforward.}, the FSI is dominated by the contribution from
$\bar qq$ pairs of transverse size $r\sim r_{FSI}$. At large $Q^2\gg
m_V^2$ and/or for production of heavy vector mesons, when
$r_{FSI}\ll R_V$, the observable $\Sigma \approx \sigma_{\bar
qq}(r_{FSI},s)$ and the nuclear transparency tend to unity from
below:
%
\BE
1 - Tr_A^{inc} \propto\,
\la T_A\ra\,
\frac{Y^2}{Q^2 + m_V^2}\,,
\label{438}
\EE
%
where $\la T_A\ra$ is the mean nuclear thickness given by
%
\BE
\la T_A\ra = \frac{\int d^2b T_A(b)^2}{A}\, .
\label{438a}
\EE
%
The expression (\ref{438}) holds for $1 - Tr_A^{inc}\ll 1$.

The nuclear modification factor $f_{inc}(s,Q^2,A)$ in
Eq.~(\ref{490}) measures the nuclear modification of the nucleon
$L/T$- ratio. Using Eq.~(\ref{438}) and different scanning
properties for the production of L and T polarized vector mesons,
one can write the following expression
%
\BE
f_{inc}(Q^2,A) - 1 \propto\,
\la T_A\ra \,
\frac{\Delta Y_{TL}^2}{Q^2 + m_V^2}\, ,
\label{439}
\EE
%
where $\Delta Y_{TL}^2$ is given by Eq.~(\ref{377BBB}). Thus at
large $Q^2\gg m_V^2$ the factor $f_{inc}$ tends to unity from above.
On the r.h.s. of Eq.~(\ref{439}) the mean nuclear thickness causes a
rise of $f_{inc}$ with $A$, whereas the fraction is responsible for
the $Q^2$ dependence.

As was already mentioned in Sect.~\ref{lcc}, different scale
parameters $Y_L$ and $Y_T$ lead to different scanning properties for
the production of L and T polarized vector mesons. In fact, the L
production amplitude is controlled by a smaller dipole size than the
T amplitude ($Y_L < Y_T$). Therefore

(i) for bottonium production $Y_T\doteq Y_L\sim 6$, the variable
$\Delta Y_{TL}^2 \to 0$, and the nuclear modification factor
$f_{inc}\sim 1$, for any fixed mass number $A$ of the nuclear
target. Consequently, the $Q^2$ dependence of the nuclear
$L/T$-ratio is almost exactly given by the analogous ratio $R_{LT}$
for the process on a nucleon target.

(ii) for charmonium production both parameters $Y_L$ and $Y_T$
slightly depend on $Q^2$, and do not differ much to each other.
Consequently, the factor $f_{inc} > 1$ does not differ much from
unity, and gradually decreases with $Q^2$ tending to unity at large
$Q^2\gg m_V^2$. According to Eq.~(\ref{439}), this deviation of
$f_{inc}$ from unity rises weakly with $A$.

\begin{figure}[bht]
\includegraphics{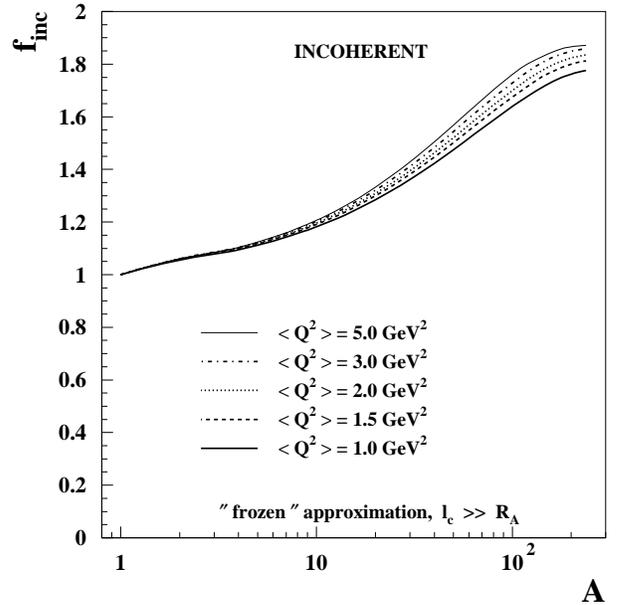}
\begin{center}
\vspace{8.0cm}
{\caption[Delta]
 {\sl $A$ dependence of the
nuclear modification factor $f_{inc} = Tr_A^{inc}(L)/Tr_A^{inc}(T)$
as the ratio of nuclear transparencies for incoherent production of
L and T polarized $\rho^0$ mesons, at different fixed values of $\la
Q^2\ra$. Calculations are performed in the limit of long coherence
length, $l_c\gg R_A$. }
 \label{trainc-lt-he}}
 \end{center}
 \end{figure}

(iii) the most interesting situation is in the production of light
vector mesons, where one should expect a much stronger nuclear
modification of the nucleon ratio $R_{LT}$ than for heavy mesons. At
small and medium values of $Q^2$, such as $r_S\gsim R_V$, there is a
strong $Q^2$ dependence of both scale parameters $Y_L$ and $Y_T$.
Moreover, the difference between $Y_T$ and $Y_L$ rises very rapidly
with $Q^2$ (see Fig.~\ref{y-lt}), resulting in a strong $Q^2$
behavior of $\Delta Y_{TL}^2$. The rise with $Q^2$ of $\Delta
Y_{TL}^2$ in the numerator of the r.h.s. of Eq.~(\ref{439}) can
fully compensate or even overcompensate a decrease of the r.h.s. of
Eq.~(\ref{439}), with $(Q^2 + m_V^2)$. This fact causes a weak
$Q^2$-rise of the nuclear modification factor. Such expectation is
confirmed by the Fig.~\ref{trainc-lt-he}, where we present the $A$
dependence of $f_{inc}$ for incoherent production of $\rho^0$
mesons, at several values of $\la Q^2\ra$ and at $\nu = 15\GeV$,
corresponding to HERMES kinematics.
%
%

According to Eq.~(\ref{439}), at fixed value of $\la Q^2\ra$ one
should expect a monotonic $A$-rise of $f_{inc}$, caused by the mean
nuclear thickness $\la T_A\ra$. This is in accordance with the
predictions presented in Fig.~\ref{trainc-lt-he}, where one can see
quite a strong nuclear modification of the nucleon $L/T$ ratio for
heavy nuclei.
%
  \begin{figure}[bht]
\includegraphics{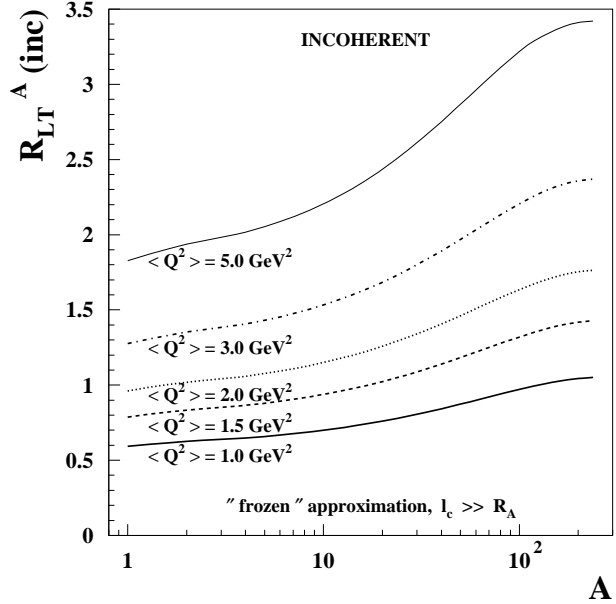}
\begin{center}
\vspace{8.0cm}
{\caption[Delta]
 {\sl $A$ dependence of the
nuclear ratio $R_{LT}^A(inc)$ (\ref{490}) of the cross sections
(\ref{510}), for incoherent production of L and T polarized $\rho^0$
mesons off nuclei, and at different fixed values of $\la Q^2\ra$.
Calculations are performed in the limit of long coherence length,
$l_c\gg R_A$. }
 \label{ltainc-he}}
 \end{center}
 \end{figure}
%

Using our results for the nucleon $L/T$-ratio (see
Fig.~\ref{r-lt-q2-n}) and for the nuclear modification factor
$f_{inc}$ (see Fig.~\ref{trainc-lt-he}), we calculated also the
nuclear $L/T$-ratio. The results are depicted in
Fig.~\ref{ltainc-he}. One can see again a monotonic $A$ dependence
of $R_{LT}^A(inc)$, coming from the $A$-behavior of $f_{inc}$.

Finally, we would like to emphasize that the discussion presented
above concerns the high energy limit $l_c\gg R_A$, when the $\bar
qq$ fluctuations can be treated as ``frozen'' during the propagation
through the nuclear target. This simplification was used for a
better and more intuitive qualitative understanding of the $Q^2$ and
$A$ behavior of the nuclear ratio $R_{LT}^{A}(inc)$. In this
``frozen'' approximation any rise of nuclear transparency with $Q^2$
represents a net manifestation of CT \cite{knst-01,n-02,n-03},
because CL effects are negligible. Generally, at smaller $l_c\lsim
R_A$, when fluctuations of the size of the $\bar qq$ pair become
important, one should include in addition also CL effects, and
therefore go beyond the simplified ``frozen'' approximation. Thus in
this kinematical region one should solve the problem of CT-CL
mixing. Both CT and CL effects are naturally incorporated in the LC
Green function formalism, and the corresponding formulae become much
more complicated, as one can see above. As was analyzed in detail in
Ref.~\cite{knst-01}, the effects of CL can mock the signal of CT if
the coherence length varies from long to short compared to the
nuclear size. In this case the nuclear transparency rises with $Q^2$
because the length of the path in nuclear matter becomes shorter,
and the vector meson (or $\bar qq$) attenuates less. Consequently,
the effects of CL lead to a stronger $Q^2$ dependence of
$Tr_A^{inc}$ than in the ``frozen'' approximation, because both
effects work in the same direction. This leads to the following
expectations:

i) According to the scanning phenomenon (\ref{10}) and
Eq.~(\ref{439}) one should expect a little bit stronger $Q^2$
dependence of $f_{inc}$. However, in the HERMES kinematical range
the formation length $l_f\gsim l_c$ and the CL $l_c\sim R_A$ and
varies with $Q^2$ approximately from $4$ to $1\fm$. Then a different
interplay of coherence and formation effects at different values of
$Q^2$ and $A$ can modify or even change the expected monotonic $Q^2$
dependence of $f_{inc}$ (see the next Sect.~\ref{incoh-data}).

ii) The monotonic $A$ dependence of $f_{inc}$ and/or $R_{LT}^A(inc)$
should remain. The CT-CL mixing can only modify the rate of the $A$-
rise of the nuclear modification factor $f_{inc}$ and/or
$R_{LT}^A(inc)$.

In conclusion we expect that the realistic calculations performed
within the LC Green function approach do not affect significantly
the expectations and conclusions concerning the $Q^2 $ and $A$
dependence of the nuclear ratio $R_{LT}^{A}(inc)$ presented above in
the ``frozen'' approximation.

%
%
\subsection{Realistic predictions for the nuclear ratio
$R_{LT}^A(inc)$}\label{incoh-data}
%
%

Exclusive incoherent electroproduction of vector mesons off nuclei
has been suggested in \cite{knnz94,knst-01} to be a very convenient
process for the investigation of CT. Increasing the photon
virtuality $Q^2$ one squeezes the produced $\bar qq$ wave packet,
and such a small colorless system propagates through the nucleus
with little attenuation, provided that the energy is sufficiently
high ($l_f\gg R_A$) so the fluctuations of the $\bar qq$ separation
are ``frozen'' during propagation. Consequently, a rise of nuclear
transparency $Tr_A^{inc}(Q^2)$ with $Q^2$ should give a signal for
CT. Indeed, such a rise was observed in the E665 experiment
\cite{e665-rho} at Fermilab for exclusive production of $\rho^0$
mesons off nuclei, and this has been claimed as a manifestation of
CT. However, the effect of coherence length \cite{kn95,hkn} leads
also to a rise of $Tr_A^{inc}(Q^2)$ with $Q^2$, and therefore
imitates the CT effects. Both effects work in the same direction and
so from this the problem of CT-CL separation arises, although this
has been already solved in refs.~\cite{hk-97,knst-01}, where a
simple prescription for the elimination of CL effects from the data
on the $Q^2$ dependence of nuclear transparency was presented. One
should bin the data in a way which keeps $l_c = const$. It means
that one should vary simultaneously $\nu$ and $Q^2$ maintaining the
CL Eq.~(\ref{30}) constant,
%
 \beq
\nu = {1\over2}\,l_c\,(Q^2+m_V^2)\ .
\label{534}
 \eeq
%
In this case any rise with $Q^2$ of nuclear transparency signals CT
\cite{hk-97,knst-01}.

In the present paper we investigate differences and peculiarities in
the production of vector mesons at different polarizations. The data
are usually presented as the ratio of the nuclear cross sections for
production of L and T polarized vector mesons. Dependence of this
ratio on various variables demonstrates different properties and
phenomena in the production of vector mesons, at separated
polarizations. Therefore it is interesting to study the $Q^2$ and
$A$ behavior of the nuclear ratio $R_{LT}^A(inc) =
\sigma_A^{inc}(L)/\sigma_A^{inc}(T)$ as a manifestation of the
polarization dependence of the CT and CL effects. Because new data
from the HERMES collaboration will appear soon we provide
predictions for the nuclear ratio $R_{LT}^{A}(inc)$ in the HERMES
kinematical range, and analyze the corresponding phenomena.
%
  \begin{figure}[bht]
\includegraphics{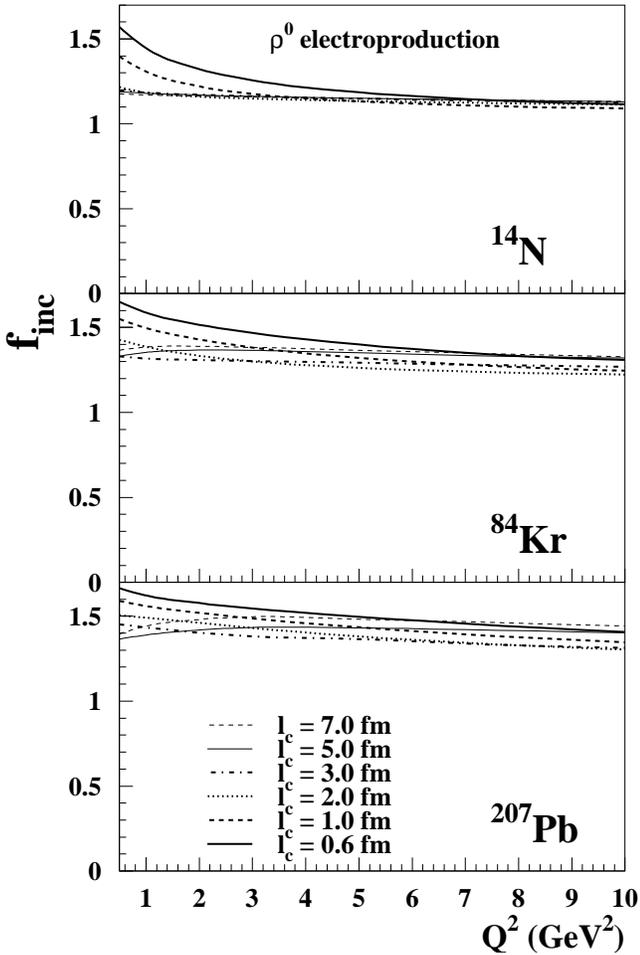}
\begin{center}
\vspace{12.7cm}
{\caption[Delta]
 {\sl $Q^2$ dependence of the ratio
$f_{inc} = Tr_A^{inc}(L)/Tr_A^{inc}(T)$ of nuclear transparencies
for incoherent production of L and T polarized $\rho^0$ mesons on
nuclear targets $^{14}N$, $^{84}Kr$ and $^{207}Pb$ (from top to
bottom). The CL Eq.~(\ref{30}) is fixed at $l_c = 0.6$, $1.0$,
$2.0$, $3.0$, $5.0$ and $7.0\fm$. }
 \label{finc-lc-all}}
 \end{center}
 \end{figure}
%

Motivated by the expected new data from the HERMES collaboration we
concentrate in the present paper on the production of light vector
mesons ($\rho^0$ and $\Phi^0$). Because the results of the
calculation for the production of $\rho^0$ and $\Phi^0$ are quite
similar we present predictions only for $\rho^0$ mesons. On the
other hand, as was discussed in Refs.~\cite{knst-01,n-02,n-03}, the
coherence and formation effects in electroproduction of vector
mesons off nuclei are much more visible for light than for heavy
vector mesons. The same happens for the study of differences in
electroproduction of L and T polarized vector mesons. The LC Green
function technique is a very effective tool for such studies because
both CT and CL effects are naturally incorporated.
%
  \begin{figure}[bht]
\includegraphics{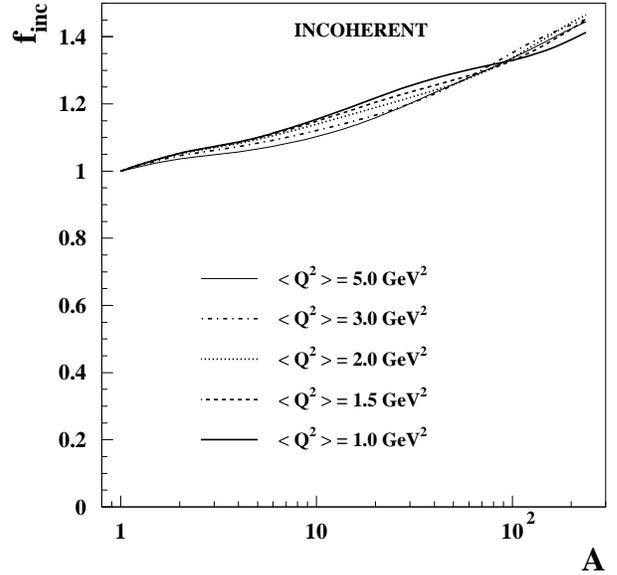}
\begin{center}
\vspace{8.0cm}
{\caption[Delta]
 {\sl $A$ dependence of the ratio
$f_{inc} = Tr_A^{inc}(L)/Tr_A^{inc}(T)$ of nuclear transparencies
for incoherent production of $L$ and $T$ polarized $\rho^0$ mesons
at different fixed values of $\la Q^2\ra$. Calculations are
performed at the photon energy $\nu = 15\GeV$. }
 \label{trainc-lt}}
 \end{center}
 \end{figure}
%

According to Eqs.~(\ref{65}) and (\ref{490}) the nuclear
modification factor $f_{inc}$ (or ratio
$Tr_A^{inc}(L)/Tr_A^{inc}(T)$ of nuclear transparencies) for
incoherent production of L and T polarized vector mesons represents
the strength of the nuclear modification of the nucleon ratio
$R_{LT}$. Therefore besides the nuclear ratio $R_{LT}^A(inc)$ the
ratio $f_{inc}$ is also a very effective variable for the study of
differences in the production of L and T polarized vector mesons off
nuclei.

First we investigate different manifestations of net CT effects in
incoherent electroproduction of L and T polarized $\rho^0$ mesons,
using the prescription (\ref{534}), which states that one should
study the $Q^2$ dependence of the factor $f_{inc}$ at fixed values
of the CL Eq.~(\ref{30}). According to the scanning phenomenon (see
Eq.~(\ref{10}) and Fig.~\ref{y-lt}), for incoherent
electroproduction of L polarized vector mesons one expects a
stronger CT effects than for T polarized vector mesons.
Consequently, at arbitrary $Q^2$ the nuclear transparency
$Tr_A^{inc}(L) > Tr_A^{inc}(T)$ and the nuclear modification factor
$f_{inc} > 1$. The results of $f_{inc}$ for incoherent production of
$\rho^0$ at several values of $l_c = 0.6$, $1.0$, $2.0$, $3.0$,
$5.0$ and $7\fm$ are presented in Fig.~\ref{finc-lc-all} for
nitrogen, krypton and lead. One can see that:

i) The nuclear modification factor decreases slightly with $Q^2$, and at
fixed $l_c$ the photon energy rises with $Q^2$. Because of a weaker
$Q^2$ dependence of the nuclear transparency at larger photon
energy, there is also a smaller difference between $Tr_A^{inc}(L)$
and $Tr_A^{inc}(T)$, i.e. a smaller value of $f_{inc}$.

ii) The $Q^2$ dependence of $f_{inc}$ is stronger at smaller $l_c$.
In fact, if the coherence length is long then the formation length is also long, 
$l_f\gsim l_c\gg R_A$, and nuclear transparency rises with
$Q^2$ only because the mean transverse separation of the $\bar qq$
fluctuation decreases. Because the production of L polarized vector
mesons is scanned at smaller $\bar qq$ transverse separations, the
nuclear transparency $Tr_A^{inc}(L) > Tr_A^{inc}(T)$ and $f_{inc} >
1$. If, however, $l_c\lsim R_A$ and fixed, the photon energy rises
with $Q^2$ and the formation length Eq.~(\ref{20}) rises as well.
Thus, these two effects, the $Q^2$ dependence of $l_f$ and the $\bar
qq$ transverse size, add up and lead to a steeper growth of
$Tr_A^{inc}(Q^2)$ for short $l_c$. Consequently, this stronger $Q^2$
dependence leads to a larger difference between $Tr_A^{inc}(L)$
and $Tr_A^{inc}(T)$, i.e. to a larger value of $f_{inc}$.

iii) The weak $Q^2$- rise of $f_{inc}$ at large $l_c\gsim 5\fm$ is
given by the Reggeon part contribution to the dipole cross section,
Eq.~(\ref{145}).

In Fig.~\ref{trainc-lt} we present the $A$ dependence of the ratio
$f_{inc}$ at $\nu = 15\GeV$ and at several fixed values of $Q^2$,
corresponding to the HERMES kinematical range. One can see that $f_{inc}
> 1$ as a consequence of the different scanning properties of
$Tr_A^{inc}(L)$ and $Tr_A^{inc}(T)$ (see Eq.~(\ref{439}) and
subsequent discussion). Notice the weak $Q^2$ dependence of $f_{inc}$,
coming from the factor $\Delta Y_{TL}^2/(Q^2 + m_V^2)$ on the r.h.s.
of Eq.~(\ref{439}). However, in contrast to the results from the
``frozen'' approximation (see Fig.~\ref{trainc-lt-he}) the nuclear
modification factor $f_{inc}$ decreases now slightly with $Q^2$ as a
consequence of a strong CT-CL mixing. Moreover, at larger values of
$A\gsim 84$ there is a change in the order of the curves calculated for
different values of $Q^2$. It is a manifestation of a different
interplay of coherence and formation effects as a function of $Q^2$
and $A$. At larger $Q^2$ the effects of CL become more important
also for lighter nuclei, when a condition $l_c\lsim R_A$ starts to be
effective.

As we already discussed in Sect.~\ref{rlt-inc}, the $A$ dependence
of the nuclear factor $f_{inc}$ comes, in the high energy limit, from
the $A$-dependent mean nuclear thickness (see Eqs.~(\ref{438}) and
(\ref{438a})). Fig.~\ref{trainc-lt} shows that performing 
realistic calculations (without restrictions on the coherence
length) we predict also a monotonic $A$-rise of $f_{inc}$, similar
to that obtained in the ``frozen'' approximation (see
Fig.~\ref{trainc-lt-he}), because both CT and CL effects work in
the same direction. However, in comparison with the ``frozen''
approximation, the $A$ dependence of the CL-CT mixing causes a
decrease of the $A$-growth of $f_{inc}$.
%
  \begin{figure}[bht]
\includegraphics{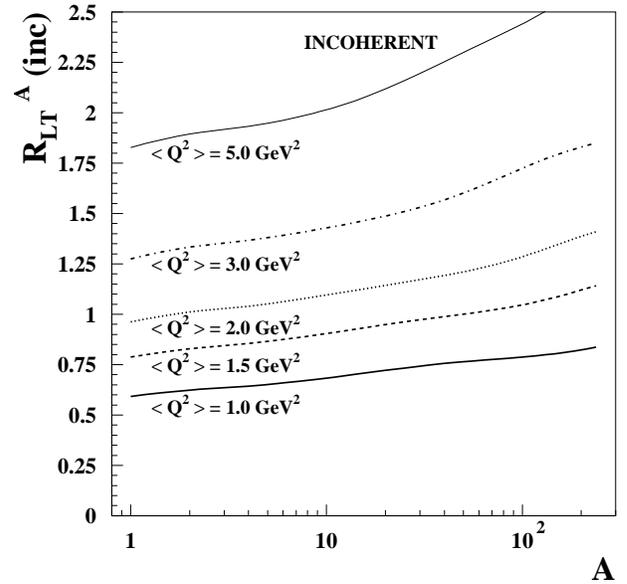}
\begin{center}
\vspace{8.0cm}
{\caption[Delta]
 {\sl $A$ dependence of the nuclear ratio
$R_{LT}^A(inc)$ (\ref{490}) of the cross sections for incoherent
production of $L$ and $T$ polarized $\rho^0$ mesons off nuclei, at
different fixed values of $\la Q^2\ra$. Calculations are performed
at the photon energy $\nu = 15\GeV$. }
 \label{ltainc}}
 \end{center}
 \end{figure}
%

According to Eq.~(\ref{490}), using known values for the nuclear
modification factor $f_{inc}$ (see Fig.~\ref{trainc-lt}) and the
nucleon $L/T$-ratio (see Fig.~\ref{r-lt-q2-n}), we present in
Fig.~\ref{ltainc} the $A$ dependence of the nuclear ratio
$R_{LT}^A(inc)$. The predictions are shown at several values of $\la
Q^2\ra$ and at $\nu = 15\,GeV$, corresponding to the HERMES kinematical
range. The $Q^2$ dependence of $R_{LT}^A(inc)$ is given by the
convolution of the $Q^2$ behavior of the nucleon ratio $R_{LT}$ (see
Fig.~\ref{r-lt-q2-n}) with nuclear factor $f_{inc}$ (see
Fig.~\ref{trainc-lt}). One can see a monotonic increase of the $A$ dependence of
$R_{LT}^A(inc)$ as a consequence of the monotonic increase with $A$ behavior of
$f_{inc}$.

%
%
\section{Coherent production of vector mesons}\label{vm-coh}
%
%

%
\subsection{The LC Green function formalism}\label{lc-coh}
%

If electroproduction of a vector meson leaves the target intact, the
process is usually called coherent or elastic, and the mesons produced
at different longitudinal coordinates and impact parameters add up
coherently. This fact considerably simplifies the expressions for
the cross sections, compared to the case of incoherent production.
The integrated cross section has the form,
%
 \BA
\sigma_A^{coh}\equiv
\sigma_{\gamma^{*}A\to VA}^{coh} &=&
\int d^2q\,\left|\int d^2b\,
e^{i\vec q\cdot\vec b}\,
{\cal M}_{\gamma^{*}A\to VA}^{coh}(b)
\right|^2
\nonumber \\
&=&
\int d^{2}\,{b}\,
|{\cal M}_{\gamma^{*}A\to VA}^{coh}
({b})\,|^{2}\ ,
\label{550}
 \EA
%
 where the coherent nuclear production amplitude is expressed as
%
 \BE
{\cal M}_{\gamma^{*}A\to VA}^{coh}({b}) =
\int\limits_{-\infty}^{\infty}\,dz\,\rho_{A}({b},z)\,
F_{1}({b},z)
\label{560}
 \EE
%
and the function $F_{1}({b},z)$ is defined by Eq.~(\ref{505}).

Contrary to incoherent vector meson production, the $t$-slopes
of the differential cross sections for nucleon and nuclear targets
are different and do not cancel in the ratio. Therefore, the
coherent nuclear transparency also includes the slope parameter
$B_{\gamma^*N}$ for the process $\gamma^{*}\,N\rightarrow V\,N$,
\footnote{ Note that in contrast to incoherent production, where
nuclear transparency is expected to saturate as $Tr^{inc}_A(Q^2) \to
1$ at large $Q^2$, for the coherent process nuclear transparency
reaches a higher limit, $Tr^{coh}_A(Q^2) \to A^{1/3}$.}
%
 \BE
Tr_{A}^{coh} = \frac{\sigma_{A}^{coh}}{A\,\sigma_{N}} =
\frac{16\,\pi\,B_{\gamma^*N}\,\sigma_{A}^{coh}}{A\,
|{\cal M}_{\gamma^{*}N\to VN}(s,Q^{2})\,|^{2}}
\label{570}
 \EE
%

Because we study the $L/T$-ratio of nuclear cross sections,
just as for incoherent vector meson production (see
Eq.~(\ref{490})) one can define the coherent nuclear ratio
$R_{LT}^{A}(coh)$ as
%
\BA
\!\!\!\!\!\!\!\!\!\!\!R_{LT}^{A}(coh)\!\!&=&\!\!
\frac{\sigma_{\gamma_{L}^*\,A\to V_L\,A}^{coh}}
     {\sigma_{\gamma_{T}^*\,A\to V_T\,A}^{coh}}
\nonumber\\
&=&\!\!
R_{LT}\,\frac{Tr_A^{coh}(L)}{Tr_A^{coh}(T)} =
R_{LT}\,f_{coh}(s,Q^2,A)\, ,
\label{617}
\EA
%
where $Tr_A^{coh}(L)$ and $Tr_A^{coh}(T)$ are defined by
Eq.~(\ref{570}) and represent the nuclear transparencies for
coherent production of L and T polarized vector mesons,
respectively. The nucleon $L/T$-ratio $R_{LT}$ in (\ref{617}) is
defined by Eq.~(\ref{377}).

%
\subsection{The nuclear ratio $R_{LT}^{A}(coh)$
in the limit
of long coherence length $l_c\gg R_A$}\label{rlt-coh}
%

Expression (\ref{550}) is
simplified in the limit of long coherence time ($l_c\gg R_A$) as
%
\begin{widetext}
 \beqn
\sigma_{A}^{coh}
\Bigr|_{l_c\gg R_A} &=&
4 \,
\int d^2 b\,\left|
\int d^2 r\,
\Biggl\{1\ -\ \exp\left[-\ {1\over2}\,
\sigma_{\bar qq}(\vec r,s)\,T_A(b)\right]\Biggr\}
\int\limits_0^1 d\alpha\,
\Psi^{*}_{V}(\vec r,\alpha)\,
\Psi_{\gamma^{*}}(\vec r,\alpha)
\right|^2\ .
\label{615}
 \eeqn
\end{widetext}
%

Here again, for the sake of clarity in the subsequent discussion we
assume the ``frozen'' approximation ($l_c\gg R_A$), which simplifies
the expressions for the cross sections and allows to understand on
a qualitative level the differences between coherent production of
L and T polarized vector mesons. The generalization of this long-$l_c$
limit to a more complicated realistic case using LC Green function
approach will be discussed below in Sect.~\ref{coh-data}.

In the limit $l_c\gg R_A$ the total integrated cross
section for coherent vector meson production
is given by Eq.~(\ref{615}), and consequently the nuclear
ratio $R_{LT}^A(coh)$ can be written as
%
\begin{widetext}
\BA
R_{LT}^A(coh)\, =\, R_{LT}\,\,\frac{B_L}
{B_T}\,\,.\,\,
\frac{\int d^2 b\, T^2_A(b)\,\biggl [1 - \frac{1}{2}\,\Sigma_L\,T_A(b) +
\ldots
\biggr ]}
     {\int d^2 b\, T^2_A(b)\,\biggl [1 - \frac{1}{2}\,\Sigma_T\,T_A(b) +
\cdots
\biggr ]}\, =\,
R_{LT}\,\frac{B_L}
{B_T}\,\,.\,\,
\frac{\la T_A\ra  - \frac{1}{2}\,\Sigma_L\,\la T_A^2\ra + \cdots}
     {\la T_A\ra  - \frac{1}{2}\,\Sigma_T\,\la T_A^2\ra + \cdots}\, ,
\label{618}
\EA
\end{widetext}
%
where the mean nuclear thickness $\la T_A\ra$ is defined by
Eq.~(\ref{438a}), and the mean nuclear thickness squared $\la
T_A^2\ra$ is given by
%
\BE
\la T_A^2\ra\, =\, \frac{\int d^2 b\, T_A(b)^3}{A}\, .
\label{619}
\EE
%
In Eq.~(\ref{618}) the variable $\Sigma$ is defined by
Eq.~(\ref{437}) and for simplicity we have explicitly shown only the
leading term of the FSI.

As we discussed in the Sect.~\ref{vm-incoh}, the FSI is dominated by the
contribution from $\bar qq$ pairs of transverse size $r\sim r_{FSI}
= 5/3 r_S$. At large $Q^2\gg m_V^2$ and/or for production of heavy
vector mesons, when $r_{FSI}\ll R_V$ the observable $\Sigma \approx
\sigma_{\bar qq}(r_{FSI},s)$, and according to the scanning phenomenon
(\ref{10}), the function $1 - g(Q^2)\,Tr_A^{coh}$ scales with $(Q^2
+ m_V^2)$ (compare with Eq.~(\ref{438})),
%
\BE
1 - g(Q^2)\,\,Tr_A^{coh} \propto
\la T_A\ra\,\,
\frac{Y^2}{Q^2 + m_V^2}\, ,
\label{620}
\EE
%
where the $Q^2$-dependent function $g(Q^2)$
reads
%
\BE
g(Q^2) = \frac{1}{\la T_A\ra}\,
\frac{1}{16\,\pi\,B(Q^2)}\,.
\label{630}
\EE
%
The relation (\ref{620}) holds for $1 - g(Q^2)\,Tr_A^{coh}\ll 1$.

It can be seen from Eq.~(\ref{617}) that, in analogy with $f_{inc}$, one
can define the coherent nuclear modification factor $f_{coh}$ as the
ratio of the coherent nuclear and nucleon $L/T$-ratio. A deviation
of $f_{coh}$ from unity as a function of $Q^2$ and $A$ provides 
information about how the coherence and formation effects manifest
themselves in coherent electroproduction of vector mesons at
different polarizations L and T. Therefore now we 
discuss the $Q^2$ and $A$ dependence of $f_{coh}$. For this purpose it
is convenient to write the following expression, using
Eqs.~(\ref{618}), (\ref{620}) and (\ref{630}),
%
\BE
\!\frac{B_{T}}
     {B_{L}}\,\,
f_{coh} - 1\,
\propto\,
\frac{\Delta Y_{TL}^2}{Q^2 + m_V^2}\,\,
\frac{\la T_A^2\ra}{\la T_A\ra}\,
\approx\,
\la T_A\ra\,
\frac{\Delta Y_{TL}^2}{Q^2 + m_V^2}\,\,
\label{640}
\EE
%
where $\la T_A^2\ra$ is defined by Eq.~(\ref{619}).
Within the discussed ``frozen'' approximation we include
for simplicity in Eq.~(\ref{640}) only
the leading term of the FSI, when $Q^2\gg m_V^2$.
%
  \begin{figure}[bht]
\includegraphics{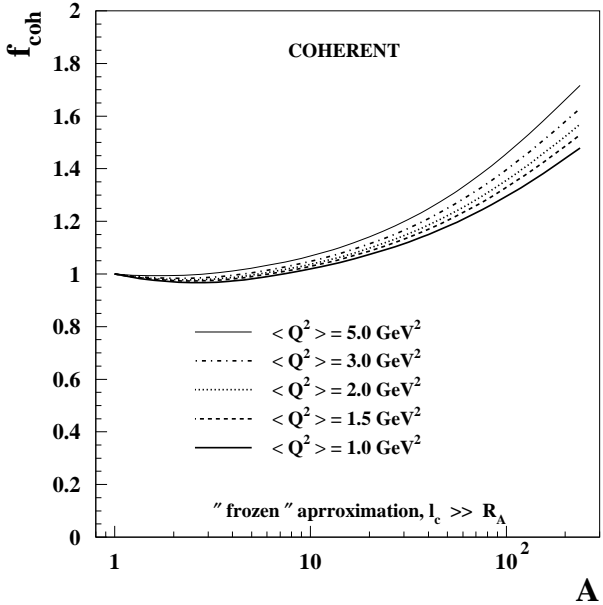}
\begin{center}
\vspace{8.0cm}
{\caption[Delta]
 {\sl $A$ dependence of the
nuclear modification factor $f_{coh} = Tr_A^{coh}(L)/Tr_A^{coh}(T)$
as the ratio of nuclear transparencies for coherent production of L
and T polarized $\rho^0$ mesons, at different fixed values of $\la
Q^2\ra$. Calculations are performed in the limit of long coherence
length, $l_c\gg R_A$. }
 \label{tracoh-lt-he}}
 \end{center}
 \end{figure}
%

Assuming the equality $B_L = B_T$, Eq.~(\ref{640}) leads to an
analogous $Q^2$ and $A$ behavior of $f_{coh}$ as the one in Eq.~(\ref{439})
for the incoherent nuclear modification factor $f_{inc}$. This is
fulfilled at large $Q^2\gg m_V^2$ and/or for the production of heavy
vector mesons, when the relativistic effects are small enough to
apply safely the nonrelativistic approximation. At small and medium
$Q^2$, however, $B_L < B_T$ \cite{nnpzz-98} and the $B_T/B_L$ ratio
on the left-hand-side (l.h.s.) of Eq.~(\ref{640}) reduces the
coherent nuclear factor $f_{coh}$. Consequently, for light nuclear
targets $A\lsim 10$ the factor $f_{coh}$ can be less than unity.

As was discussed in detail in Sect.~\ref{rlt-inc}, for bottonium
production $Y_T\doteq Y_L\sim 6$, and the slope parameters $B_L
\approx B_T$. Consequently, $f_{coh}\sim 1$ and the $Q^2$ dependence
of the nuclear $L/T$-ratio for coherent reactions is almost exactly
given by the analogous ratio $R_{LT}$ for the process on a nucleon.
This conclusion is essentially the same as the one expected
for incoherent production of bottonia.

For charmonium production both $Y_L$ and $Y_T$ depend slightly on
$Q^2$, and do not differ much from each other. Consequently, the
difference $\Delta Y_{TL}^2$ acquires a small value and rises very
weakly with $Q^2$. Because $B_T/B_L \approx 1.03$ in the
photoproduction limit, the nuclear factor $f_{coh}$ can
go below unity at small values of $Q^2$ and $A$. Increasing
$Q^2$ the ratio $B_T/B_L$ tends to unity from above and $f_{coh}$
gradually comes to unity from below at small $A$ or from above at
medium and large $A$.

In contrast to the production of heavy vector mesons, for the production
of light vector mesons we expect much larger nuclear modifications of
the nucleon ratio $R_{LT}$, just as for the incoherent processes
discussed in the previous Sect.~\ref{vm-incoh}. At small and medium
$Q^2$ such as $r_S\gsim R_V$, there is a strong $Q^2$ dependence of
$\Delta Y_{TL}^2$, which can even overcompensate the rise of $(Q^2 +
m_V^2)$ in the denominator of Eq.~(\ref{640}). This fact can
lead to a weak rise with $Q^2$ of the coherent nuclear factor
$f_{coh}$, which  is further enhanced by the decrease of
the $B_T/B_L$ ratio on the l.h.s. of Eq.~(\ref{640}). As a result,
we expect a stronger $Q^2$ dependence of $f_{coh}$ than of $f_{inc}$.
Such an expectation is supported by calculations performed in the
limit of long coherence length and is shown in Fig.~\ref{tracoh-lt-he}
(compare with Fig.~\ref{trainc-lt-he}).
%
  \begin{figure}[bht]
\includegraphics{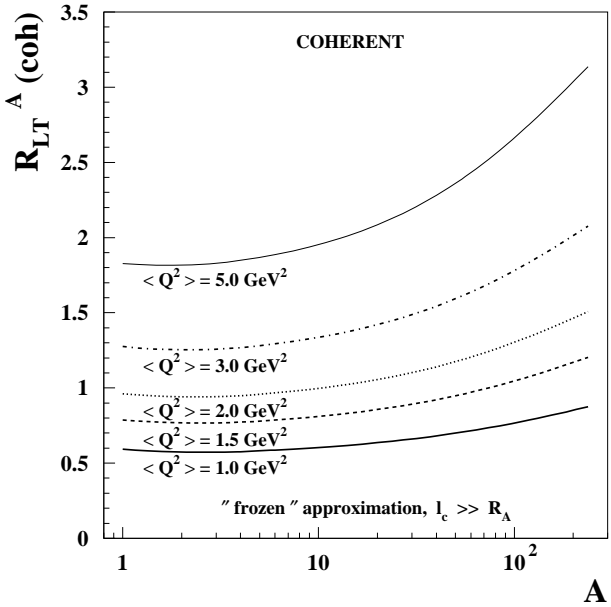}
\begin{center}
\vspace{8.0cm}
{\caption[Delta]
 {\sl $A$ dependence of the
nuclear ratio $R_{LT}^A(coh)$ (\ref{617}) of the cross sections
(\ref{615}) for coherent production of L and T polarized $\rho^0$
mesons off nuclei, at different fixed values of $\la Q^2\ra$.
Calculations are performed in the limit of long coherence length,
$l_c\gg R_A$. }
 \label{ltacoh-he}}
 \end{center}
 \end{figure}
%

Concluding, in the HERMES kinematical range, $\sim 1 < Q^2 <
5\GeV^2$, studied in the present paper, we expect a rise with $Q^2$
of the nuclear modification factor $f_{coh}$. The rate of this rise
is then given by the mean nuclear thickness, as follows from
Eq.~(\ref{640}). Consequently, we expect a monotonic rise of
$f_{coh}$ with $A$, just as for the incoherent nuclear modification
factor $f_{inc}$ (see also Fig.~\ref{trainc-lt-he}). Monotonic $A$-increase
behavior of $f_{coh}$ is confirmed also by the predictions depicted
in Fig.~\ref{tracoh-lt-he} at several values of $\la Q^2\ra$,
corresponding to the HERMES kinematical range.

For completeness we calculated also the nuclear $L/T$-ratio using
the known nuclear modification factor $f_{coh}$ and the nucleon $L/T$-ratio. 
The results are presented in Fig.~\ref{ltacoh-he}. One can
see a monotonic $A$ dependence of $R_{LT}^A(coh)$ as a consequence
of a corresponding monotonic $A$-increase behavior of $f_{coh}$.

In the following Section we demonstrate, however, that in contrast
to incoherent vector meson production such a picture of $Q^2$ and
$A$ behavior for $f_{coh}$ and/or $R_{LT}^A(coh)$ drastically
changes going beyond this ``frozen'' approximation. This is the
crucial point which leads to interesting physics in the investigation
of light vector mesons produced coherently off nuclei.

%
\subsection{Realistic predictions for the nuclear ratio $R_{LT}^A(coh)$}
\label{coh-data}
%
%
%
  \begin{figure}[bht]
\includegraphics{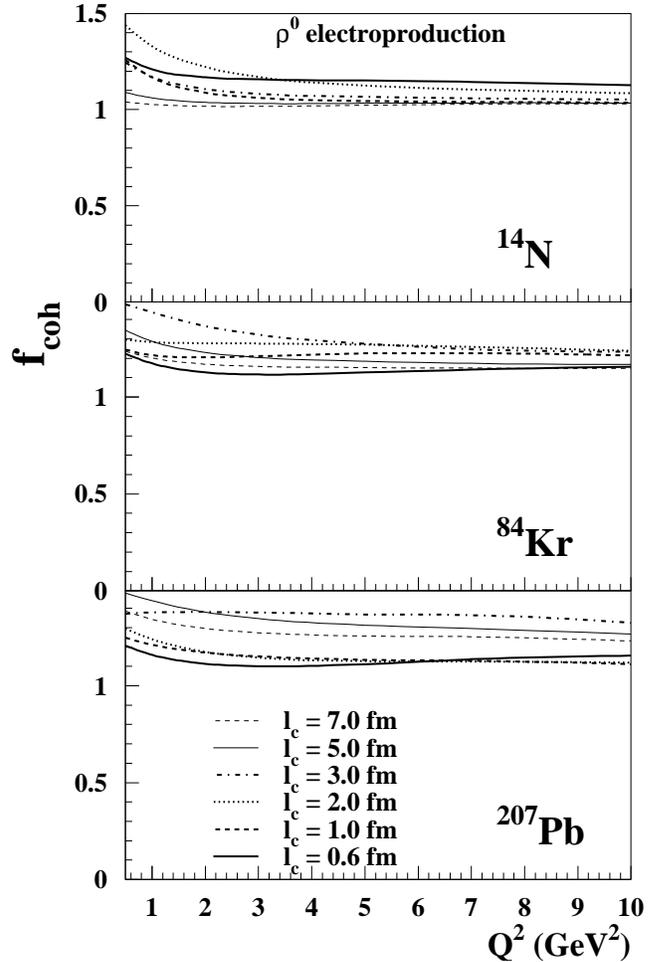}
\begin{center}
\vspace{12.7cm}
{\caption[Delta]
 {\sl $Q^2$ dependence of the ratio
$f_{coh} = Tr_A^{coh}(L)/Tr_A^{coh}(T)$
of nuclear transparencies for coherent production of L and T
polarized $\rho^0$ mesons
on nuclear targets $^{14}N$, $^{84}Kr$ and $^{207}Pb$
(from top to bottom).
The CL Eq.~(\ref{30}) is fixed at $l_c = 0.6$, $1.0$, $2.0$,
$3.0$, $5.0$ and $7.0\fm$.
}
 \label{fcoh-lc-all}}
 \end{center}
 \end{figure}
%

Analogously as was done for incoherent production of vector mesons,
here we study the differences in coherent electroproduction of L and T
polarized vector mesons off nuclei, performing a realistic
calculation without restrictions on the CL. We focus on the
production of $\rho^0$ mesons, where CT and CL effects are the most
visible. This is also supported by our expectations about the new
data from the HERMES collaboration, and therefore our calculations cover the
corresponding kinematical range. We use the LC Green
function formalism which naturally incorporates both CT and CL effects.

First we study the net CT effect in L
and T polarizations, by eliminating the effects of CL in a
way similar to what was suggested for incoherent reactions, which is
selecting experimental data with $l_c = const.$ We calculated the
nuclear modification factor $f_{coh}$ for the coherent reaction
$\gamma^*\,A\to \rho^0\,A$ as a function of $Q^2$, at different fixed
values of $l_c$. The results for $l_c = 0.6$, $1.0$, $2.0$, $3.0$,
$5.0$ and $7.0\fm$ are depicted in Fig.~\ref{fcoh-lc-all}. In
contrast to the incoherent processes one can observe a much more complicated
$Q^2$ behavior, which is the result of an interplay between CT and CL
effects when a contraction of the CL causes an effect opposite to
CT. This CL-CT mixing as a function of $Q^2$ changes the order of
curves calculated at different values of $l_c$.
%
  \begin{figure}[bht]
\includegraphics{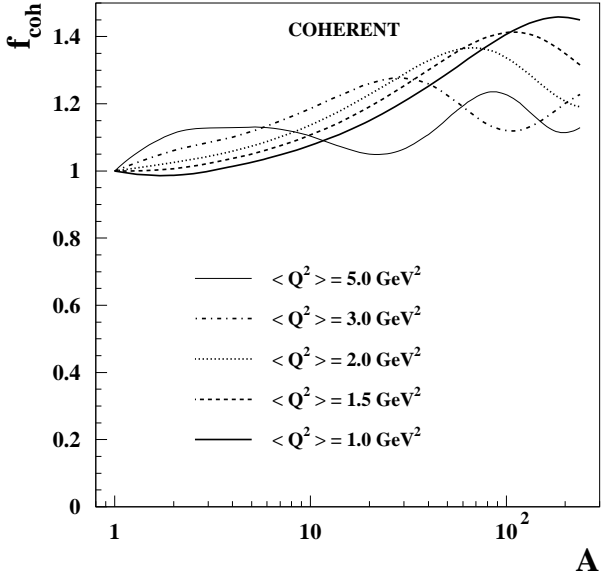}
\begin{center}
\vspace{8.0cm}
{\caption[Delta]
 {\sl $A$ dependence of the ratio
$f_{coh} = Tr_A^{coh}(L)/Tr_A^{coh}(T)$
of nuclear transparencies for coherent production of $L$ and $T$
polarized $\rho^0$ mesons at different fixed values
of $\la Q^2\ra$. Calculations are performed at the photon energy
$\nu = 15\GeV$.
}
 \label{tracoh-lt}}
 \end{center}
 \end{figure}
%

Following Eq.~(\ref{617}) the coherent nuclear factor $f_{coh}$ is
given as the ratio $Tr_A^{coh}(L)/Tr_A^{coh}(T)$ of nuclear
transparencies for coherent production of L and T polarized vector
mesons. It represents the strength of the nuclear modification of the
nucleon ratio $R_{LT}$. In Fig.~\ref{tracoh-lt} we present the $A$
dependence of $f_{coh}$ for $\rho^0$ production, at the photon energy
$\nu = 15\GeV$ and several fixed values of $\la Q^2\ra$,
corresponding to the HERMES kinematical range. One can see that the
predictions dramatically changed from those that we found for
the limit of long CL, $l_c\gg R_A$, in the previous
Sect.~\ref{rlt-coh}. The reason is that the HERMES kinematics does
not allow to neglect the effects of CL. Only at very small $A\lsim
4$ and at $Q^2\lsim 3\GeV^2$ one can assume small CL effects,
because $l_c > R_A$. Then the $A$ and $Q^2$ behavior of $f_{coh}$
follows the scenario described within the ``frozen'' approximation
(see Sect.~\ref{rlt-coh}), which means that $f_{coh}$ rises with $Q^2$
and has a monotonic $A$ dependence.

In \cite{knst-01} it was demonstrated that for coherent production of
vector mesons the contraction of the CL with $Q^2$ causes an effect
opposite to CT. Nuclear transparency is suppressed rather than
enhanced. At large $Q^2$ ,when $l_c\lsim R_A$, and at medium energies,
corresponding to the HERMES kinematics, the suppression of nuclear
transparency can be so strong that fully compensates or even
overcompensates the rise of nuclear transparency with $Q^2$ given by
CT. Because $Tr_A^{coh}(L)$ is scanned at smaller dipole sizes than
$Tr_A^{coh}(T)$, one can expect that at fixed $Q^2$ the former
nuclear transparency has stronger CL effects than the latter one.
This different manifestation of CL effects for L and T polarizations
depends also on $A$. Consequently, one may expect a nontrivial and
nonmonotonic $A$ and $Q^2$ dependence of the nuclear modification
factor $f_{coh}$.

Mainly due to the effects of CL, there is an unusual order of curves
at different values of $\la Q^2\ra$, as is shown in
Fig.~\ref{tracoh-lt}. Moreover, the order of curves is changed at
various values of $A$, as a consequence of the fact that the condition
$l_c\gg R_A$ is broken in a different degree for different nuclear
targets. At $\la Q^2\ra = 1\GeV^2$ the effects of CL start to be
important at $A\gsim 100$ and lead to a diminishing of the $A$-rise
of the nuclear factor $f_{coh}$. One can see by the thick solid line in Fig.~\ref{tracoh-lt}
that there is even a maximum of $f_{coh}$ at
$A\sim 200$, as a natural demonstration of the effectiveness of CL
effects. Larger $\la Q^2\ra$ leads to a contraction of the CL.
Consequently, the effect of CL-contraction becomes also important
for lighter nuclear targets, which means that the maximum is shifted to
smaller values of $A$. The combination of the $A$-rise of $f_{coh}$ through
the nuclear profile function (\ref{60}), together with the different
manifestation of CL effects as a function of $Q^2$ and $A$, lead to a
nontrivial and nonmonotonic $A$ dependence of $f_{coh}$, as is shown
in Fig.~\ref{tracoh-lt}. This $A$ dependence is even more
complicated at larger values of $\la Q^2\ra$, when the CL effects are
effective at a different level, for a broader range of nuclear
targets.
%
  \begin{figure}[bht]
\includegraphics{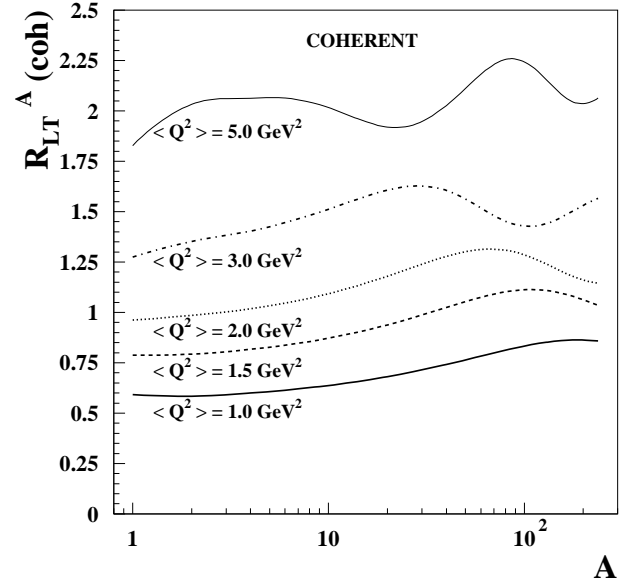}
\begin{center}
\vspace{8.0cm}
{\caption[Delta]
 {\sl $A$ dependence of the nuclear ratio
$R_{LT}^A(coh)$ (\ref{617}) of the cross sections for coherent
production of $L$ and $T$ polarized $\rho^0$ mesons off nuclei at
different fixed values of $\la Q^2\ra$. Calculations are performed
at the photon energy $\nu = 15\GeV$. }
 \label{ltacoh}}
 \end{center}
 \end{figure}
%

In Fig.~\ref{ltacoh} we present the $A$ dependence of the nuclear ratio
$R_{LT}^A(coh)$, obtained from the nuclear modification factor
$f_{coh}$  and the nucleon $L/T$-ratio (see Eq.~(\ref{617})). The
predictions are shown at several values of $\la Q^2\ra$, and at $\nu
= 15\GeV$, corresponding to the HERMES kinematical range. One can see
that a nonmonotonic $A$ dependence of $f_{coh}$ is projected into a
nonmonotonic $A$ dependence of $R_{LT}^A(coh)$. The $Q^2$ dependence
of $R_{LT}^A(coh)$ is given by the convolution of the $Q^2$ behavior
of the nucleon ratio $R_{LT}$ (see Fig.~\ref{r-lt-q2-n}) with
the nuclear factor $f_{coh}$ (see Fig.~\ref{tracoh-lt}). The predicted
anomalous $A$ behavior of the coherent nuclear ratio $R_{LT}^A(coh)$
at different values of $\la Q^2\ra$ as a undeniable and irrefutable
manifestation of strong CL effects, and can be tested by the HERMES
collaboration or at JLab.

%
\section{Summary and conclusions}\label{conclusions}
%

Electroproduction of vector mesons off nuclei is very effective tool
for the study of the interplay between coherence (shadowing) and
formation (color transparency) effects. In the present paper we
investigated how these effects manifest themselves differently in
the production of L and T polarized vector mesons off nuclei. The data
are usually presented as the $L/T$-ratio of the nuclear production
cross sections. Then an investigation of the behavior of this ratio as
a function of various variables ($Q^2$, $A$, etc.), and a deviation
of this ratio from unity, allows to study different properties and
manifestations of corresponding phenomena in the production of vector
mesons, at separated polarizations. We used, from \cite{knst-01}, a
rigorous quantum-mechanical approach based on the light-cone QCD
Green function formalism, which naturally incorporates these
interference effects. We focused on the production of
light vector mesons, because here the polarization dependence of CT and CL
effects is much more visible than in the production of heavy vector
mesons. Because new data from the HERMES collaboration are expected
to appear soon, we presented predictions for the nuclear $L/T$ ratios
(see Eqs.~(\ref{490}) and (\ref{617})) within the corresponding
kinematical range. These predictions are  made for $\rho^0$
mesons produced both coherently and incoherently off nuclei.

The strength of the nuclear modification of the nucleon $L/T$ ratio
(\ref{377}) is given by the nuclear modification factors $f_{inc}$
and $f_{coh}$, which are defined as the ratio of nuclear
transparencies for electroproduction of L and T polarized vector
mesons (see Eqs.~(\ref{490}) and (\ref{617})). If these factors are
equal to unity there are no nuclear effects. Therefore besides the nuclear
$L/T$ ratios, the nuclear modification factors are also very
effective variables for the study of differences in the production of L and
T polarized vector mesons off nuclei. The nuclear $L/T$ ratio is
then given as the product of the nucleon $L/T$ ratio and the
nuclear modification factor.

As the first step we compare the nucleon $L/T$ ratio as a function
of $Q^2$ with available data on electroproduction of $\rho^0$,
$\Phi^0$ mesons and charmonia and find a nice agreement (see
Figs.~\ref{r-lt-q2-n}, \ref{r-lt-q2-n1} and \ref{r-lt-q2-n2}). This
is a very important achievement because the nucleon $L/T$ ratio
represents a basis for the correct determination of the nuclear
$L/T$ ratio via the nuclear modification factor.

In order to obtain more intuitive information about the $A$ and $Q^2$
behavior of the nuclear $L/T$ ratio and/or nuclear modification
factor we presented on the qualitative level, using the scanning phenomenon
(\ref{10}), the corresponding predictions in the high energy limit
($l_c\gg R_A$). Here the expressions for nuclear production cross
sections are sufficiently simplified. This so called ``frozen''
approximation includes only CT because there are no fluctuations of
the transverse size of the $\bar qq$ pair. For incoherent
electroproduction of $\rho^0$ mesons we predict a very weak $Q^2$-
growth of $f_{inc}$ in the HERMES kinematical range (see
Fig.~\ref{trainc-lt-he}) due to a strong $Q^2$-rise of a difference
between the scanning radii corresponding to T and L polarizations
(see Fig.~\ref{y-lt} and Eq.~(\ref{10})). In contrast to incoherent
processes, for $\rho^0$ mesons produced coherently off nuclei one
should include in $f_{coh}$ also the slope parameters $B_L$ and
$B_T$ for different polarizations L and T (see Eq.~(\ref{640})).
Consequently, we expect a stronger $Q^2$ dependence of $f_{coh}$
(see Fig.~\ref{tracoh-lt-he}) due to different $Q^2$ dependences of
the corresponding slope parameters. We predict a monotonic rise with
$A$ of both nuclear factors $f_{inc}$ and $f_{coh}$, which comes from
the $A$- dependent mean nuclear thickness.

The ``frozen'' approximation cannot be applied for the
study of differences in electroproduction of vector mesons off
nuclei at different polarizations, in the HERMES kinematical range.
Therefore we use the approach of Ref.~\cite{knst-01}, which
interpolates between the previously known low and high energy limits
for incoherent production (see Eq.~(\ref{520})). Eq.~(\ref{560})
does the same for coherent production.

In the incoherent electroproduction of vector mesons at low and
medium energies, the onset of coherence effects (shadowing) can mimic
the expected signal of CT. Both effects, CT and CL, work in the same
direction. In comparison with the high energy limit, the onset of CL does
not change much the $A$ and $Q^2$ behavior of $f_{inc}$.
Consequently, we predict again a weak $Q^2$ dependence of $f_{inc}$
(see Fig.~\ref{trainc-lt}). Investigating the $A$ dependence of
$f_{inc}$, the interplay between CT and CL effects changes the order
of curves calculated at different values of $Q^2$. The CL-CT mixing
modifies also the rate of the $A$-rise of $f_{inc}$, but conserves
the monotonic $A$ dependence typical for the ``frozen''
approximation (see Fig.~\ref{trainc-lt}). Therefore we predict also
a monotonic $A$ increase behavior of the nuclear $L/T$ ratio at different
values of $Q^2$ (see Fig.~\ref{ltainc}).

In coherent production of vector mesons the natural incorporation of the
CL effects in the Green function formalism changes drastically the
$A$ and $Q^2$ behavior of $f_{coh}$, predicted for the high energy
limit. The contraction of the CL with $Q^2$ causes an effect
opposite to CT. There is a different manifestation of CL effects at
various values of $Q^2$ and $A$, which together with CT effects leads
to a nontrivial and anomalous $A$ and $Q^2$ dependence of the
nuclear modification factor. The nonmonotonic $A$ dependence is even
more complicated at larger values of $Q^2$ as a result of stronger
CL effects for a broader range of nuclear targets (see
Fig.~\ref{trainc-lt}). Consequently, we predict also a nonmonotonic
and anomalous $A$ dependence of the nuclear $L/T$ ratio at different
values of $Q^2$ (see Fig.~\ref{ltacoh}), which gives a motivation to
detect such anomalous manifestations of strong CL effects in
experiments with the HERMES spectrometer and especially at JLab.

We investigated also different manifestations of  net CT effects
at different polarizations L and T, using a prescription from
Refs.~\cite{hk-97,knst-01}, calculating the nuclear modification
factor as a function of $Q^2$ at various fixed values of the
coherence length. 

i) In incoherent production of $\rho^0$ mesons, where we found a stronger CT
effects for L than for T polarization, i.e. $f_{inc} > 1$. Moreover,
$f_{inc}$ rises towards small values of $Q^2$ at short $l_c\lsim
R_A$ (see Fig.~\ref{finc-lc-all}). The two effects, that is the $Q^2$
dependence of $l_f$ and the $\bar qq$ transverse size, which add up
and lead to a steeper $Q^2$-growth of nuclear transparency, and
consequently to larger values of $f_{inc}$. 

ii) In coherent production of $\rho^0$ mesons we predicted also $f_{coh} >
1$ (see Fig.~\ref{fcoh-lc-all}). However, the $Q^2$ behavior of
$f_{coh}$ is more complicated in comparison with the incoherent
reaction, which follows from the fact that a contraction of the CL with
$Q^2$ causes an effect opposite to CT. Then the CT-CL mixing as a
function of $Q^2$ changes the order of curves calculated at
different values of $l_c$.

In conclusion, the exploratory study of the $A$ dependence of the nuclear
$L/T$ ratio, especially in coherent electroproduction of light vector
mesons off nuclei, opens new possibilities for the search for the CL
effects and their different manifestations at different
polarizations with medium energy electrons.

\begin{acknowledgments}
 This work was supported in part by Fondecyt (Chile) grant 1050519,
by DFG (Germany)  grant PI182/3-1, and by the Slovak Funding
Agency, Grant No. 2/7058/27.
\end{acknowledgments}

 \def\appendix{\par
 \setcounter{section}{0} \setcounter{subsection}{0}
 \def\thesection{Appendix \Alph{section}}
 \def\thesubsection{\Alph{section}.\arabic{subsection}}
 \def\theequation{\Alph{section}.\arabic{equation}}
 \setcounter{equation}{0}}

\end{document}